\documentclass[11pt]{article}
\usepackage[utf8]{inputenc}
\usepackage[toc, page]{appendix}
\usepackage{graphicx}
\usepackage{amsmath}
\usepackage{lscape}
\usepackage{hhline}
\usepackage{authblk}
\usepackage{setspace}
\usepackage{caption}
\usepackage{hyperref}

\newcommand{\no}{\nonumber}
\newcommand{\bb}{\boldsymbol}

\def\mysingleq#1{`#1'}

\providecommand{\keywords}[1]
{
  \small	
  \textbf{\textit{Keywords:}} #1
}

\title{Seismic-phase detection using multiple deep learning models for global and local representations of waveforms}

\author[1]{Tomoki Tokuda}
\author[1, 2, *]{Hiromichi Nagao}
\affil[1]{Earthquake Research Institute, The University of Tokyo, Japan}
\affil[2]{Graduate School of Information Science and Technology, The University of Tokyo, Japan}
\affil[*]{Corresponding author: Hiromichi Nagao, nagaoh@eri.u-tokyo.ac.jp}

\date{}

\begin{document}

\maketitle

\begin{abstract}
The detection of earthquakes is a fundamental prerequisite for seismology and contributes to various research areas, such as forecasting earthquakes and understanding the crust/mantle structure. 
Recent advances in machine learning technologies have enabled the automatic detection of earthquakes from waveform data.  In particular, various state-of-the-art deep-learning methods have been applied to this endeavour. In this study, we proposed and tested a novel phase detection method employing deep learning, which is based on a standard convolutional neural network in a new framework. The novelty of the proposed method is its separate explicit learning strategy for global and local representations of waveforms, which enhances its robustness and flexibility. Prior to modelling the proposed method, we identified local representations of the waveform by the multiple clustering of waveforms, in which the data points were optimally partitioned. Based on this result, we considered a global representation and two local representations of the waveform. Subsequently, different phase detection models were trained for each global and local representation. For a new waveform, the overall phase probability was evaluated as a product of the phase probabilities of each model. This additional information on local representations makes the proposed method robust to noise, which is demonstrated by its application to the test data. Furthermore, an application to seismic swarm data demonstrated the robust performance of the proposed method compared with those of other deep learning methods. Finally, in an application to low-frequency earthquakes, we demonstrated the flexibility of the proposed method, which is readily adaptable for the detection of low-frequency earthquakes by retraining only a local model.
\end{abstract}

\keywords{Phase detection; Deep learning; Neural networks; Time-series analysis}

\section{Introduction}
The detection of earthquakes is a fundamental prerequisite for seismology and contributes to various research areas, such as forecasting earthquakes and understanding the crust/mantle structure. Using highly sensitive seismometers installed throughout the world in various locations, modern seismology allows us to detect large  earthquakes and infinitesimal ones that may be imperceptible in normal circumstances. 

Several methods that focus on human-selected characteristic features have  been proposed for earthquake detection using seismic waveform data such as amplitude and frequency \cite{zhu2022end}. A classic method is based on short-term amplitude (STA)  over long-term amplitude (LTA) averages of waveforms (STA/LTA method)\cite{stevenson1976microearthquakes,allen1978automatic, baer1987automatic}, which detects the onset of an earthquake as a sudden change in the STA/LTA ratio. Combining the STA/LTA method with an autoregressive model enables the effective inference of the P-wave's (Primary wave) arrival time based on the Akaike information criterion (AR-AIC) method \cite{baer1987automatic}. Furthermore, frequency-based methods \cite{lomax2012automatic, mousavi2016fast} detect earthquakes by focusing on specific frequency domains that characterise the dominant frequency of earthquakes. 

Recently, deep learning-based methods have gained much attention for earthquake detection \cite{mousavi2022deep}. Instead of human-selected characteristic features, a deep learning model learns specific waveform features in a data-driven manner without prior knowledge. First, using training data consisting of waveforms and seismic phase labels, the model parameters for a neural network are optimised. Subsequently, based on the trained model, earthquakes are detected for a new instance of the waveform. Several methods have been proposed based on different forward neural network architectures, such as the convolutional neural network (CNN) \cite{ross2018generalized, perol2018convolutional, yang2021simultaneous} and U-net\cite{zhu2019phasenet, woollam2019convolutional}, which have been widely used in fields other than earthquake detection, such as image classification and image segmentation. Furthermore, methods based on recurrent neural networks have been proposed \cite{zhou2019hybrid, mousavi2020earthquake, soto2021deepphasepick}. These methods aim to capture contextual information by preserving the time-sequential memory of a waveform. In particular, the method proposed in \cite{mousavi2020earthquake} incorporates the attention mechanism \cite{bahdanau2014neural} into the recurrent neural network, which allows for effective feature extraction. Moreover, a hybrid method that involves U-net and an attention mechanism has been proposed \cite{liao2022toward}. A recent review paper \cite{munchmeyer2022picker} on various deep learning-based methods suggests superior performance in seismic-phase detection using the generalised phase detection (GPD) method \cite{ross2018generalized}, PhaseNet \cite{zhu2019phasenet}, and earthquake transformer (EQT) \cite{mousavi2020earthquake}.

Note that the deep learning approach typically does not reveal the relevant features of the waveform for earthquake detection. This is a general problem of the deep learning approach, which is widely known as a \mysingleq{black box} problem \cite{rudin2019stop}. The attention mechanism \cite{yang2016hierarchical, vaswani2017attention} is a promising bail-out strategy for this problem. Implicitly, the attention mechanism
focuses on a specific part of the data, effectively learning relevant features and making it explainable for those extracted features. For earthquake detection, EQT adopts this approach, which, for seismic detection, focuses on specific parts of the waveform in a data-driven manner \cite{mousavi2020earthquake}. 

\begin{figure}
\includegraphics[scale=0.4, trim=30mm 40mm 0mm 50mm]{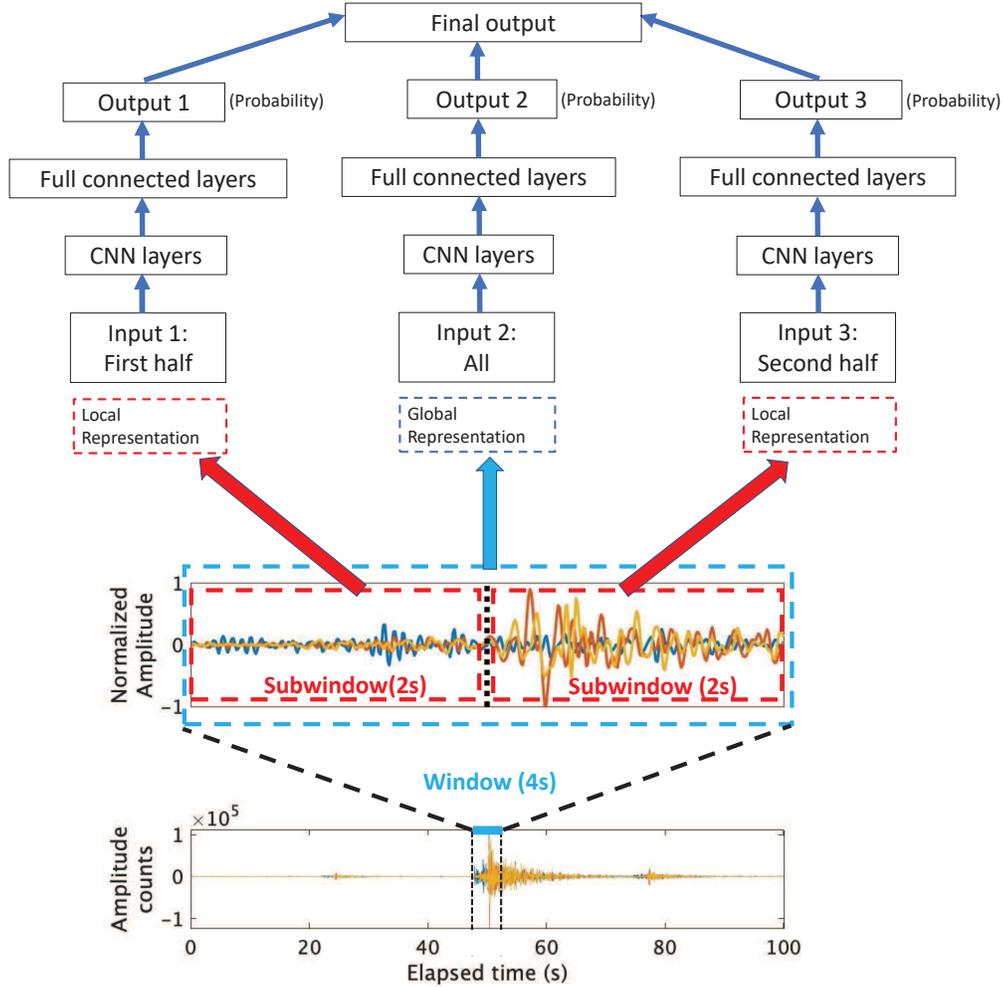}
\caption{Architecture of the proposed method. We focus on a time window (4 s) in the 3-component seismic waveform. The waveform in the time window is normalised by the maximum absolute amplitude of the waveform, which constitutes a global representation. Furthermore,
we consider sub-windows (2 s) for the first and second halves of the data points, which constitute local representations. The waveform data in each window is passed to a convolutional neural network as an input (inputs 1-3 in the illustration), and the phase probability for the P-phase, S-phase, and noise is yielded as an output (outputs 1-3). Importantly, each neural network independently undergoes supervised training. 
For a test instance of the waveform, outputs 1-3 are combined as a probabilistic product, which yields a final phase probability. Note that outputs 1-3 are the same for a training instance, whereas for a test instance, these are generally different.}
\label{architecture}
\end{figure}

In this paper, we report on a novel method for phase detection of Primary wave (P-phase), Secondary wave (S-phase) and noise, explicitly focusing on specific parts of the waveform. In contrast to EQT, we predetermine the focal parts of the waveform. We consider both global and local representations of the waveform for a 4-s time window. For global representation, we focus on the whole waveform, whereas for local representations, we focus on the first and second half of the time window (Fig.~\ref{architecture}). For each representation, we independently train a detection model using a CNN deep-learning architecture similar to the GPD model. For a new instance of a waveform, we combine the outputs of these different models in the form of a product of phase probabilities. This separate learning strategy adds more information on the features of the waveform, which makes our method more robust for earthquake detection and flexible in feature extraction.  To  our knowledge, for seismic-phase detection, there has been no neural network method that similarly uses a separate learning strategy. 

The strategy of using both global and local representations of data is closely related to context modelling for object detection in the deep learning literature \cite{liu2020deep}. The fundamental idea of context modelling is that a physical object coexists with the surrounding background and other objects; hence, context plays an important role in object detection. Conventional CNN-based deep learning is supposed to capture contextual information implicitly with multiple levels of abstraction \cite{liu2020deep}, but it can potentially overlook local details \cite{farabet2012learning, zeng2016gated}. Therefore, it is beneficial to explicitly model a network structure that considers both global and local contexts. The multi-region CNN model \cite{gidaris2015object} is a pioneering work that uses this strategy to extract features from several  different regions, such as half regions, border regions, and central regions. These extracted features were combined in the final layer of the neural network for object detection. Moreover, several variants related to this method have been proposed. The gated bidirectional CNN method \cite{zeng2016gated} allows interactions between multiscale context regions in feature extraction. In doing so, local contexts are expected to complement each other in validating the CNN’s feature extraction. The attention to context CNN \cite{li2017scene, zhu2017couplenet} method considers different architectures for global and local contexts, which allows for the effective capture of contextual locations.  

In the following sections, we first introduce the concepts of global and local waveform representations. Local representations were identified in a data-driven manner using multiple clustering analyses. Subsequently, we developed our method, in which the phase probability is defined by the product of the phase probabilities of these representation models. Next, we show that the proposed method outperforms the GPD model when the data are contaminated by noise. For application to continuous seismic waveform data, we demonstrate the high performance of our method compared with the GPD model, PhaseNet, and EQT on the 2016 Bombay Beach  swarm.  Furthermore, we show that our method can be adapted for the detection of low-frequency earthquakes (LFEs) by retraining a local representation model without a real LFE waveform. These results imply the robustness and flexibility of our proposed method, which makes the best use of global and local information on waveform by a separate learning strategy for different focal parts. 

\section{Global and local representations}\label{intro2}
\begin{figure}
\includegraphics[scale=0.7, trim=0mm 0mm 0mm 20mm]{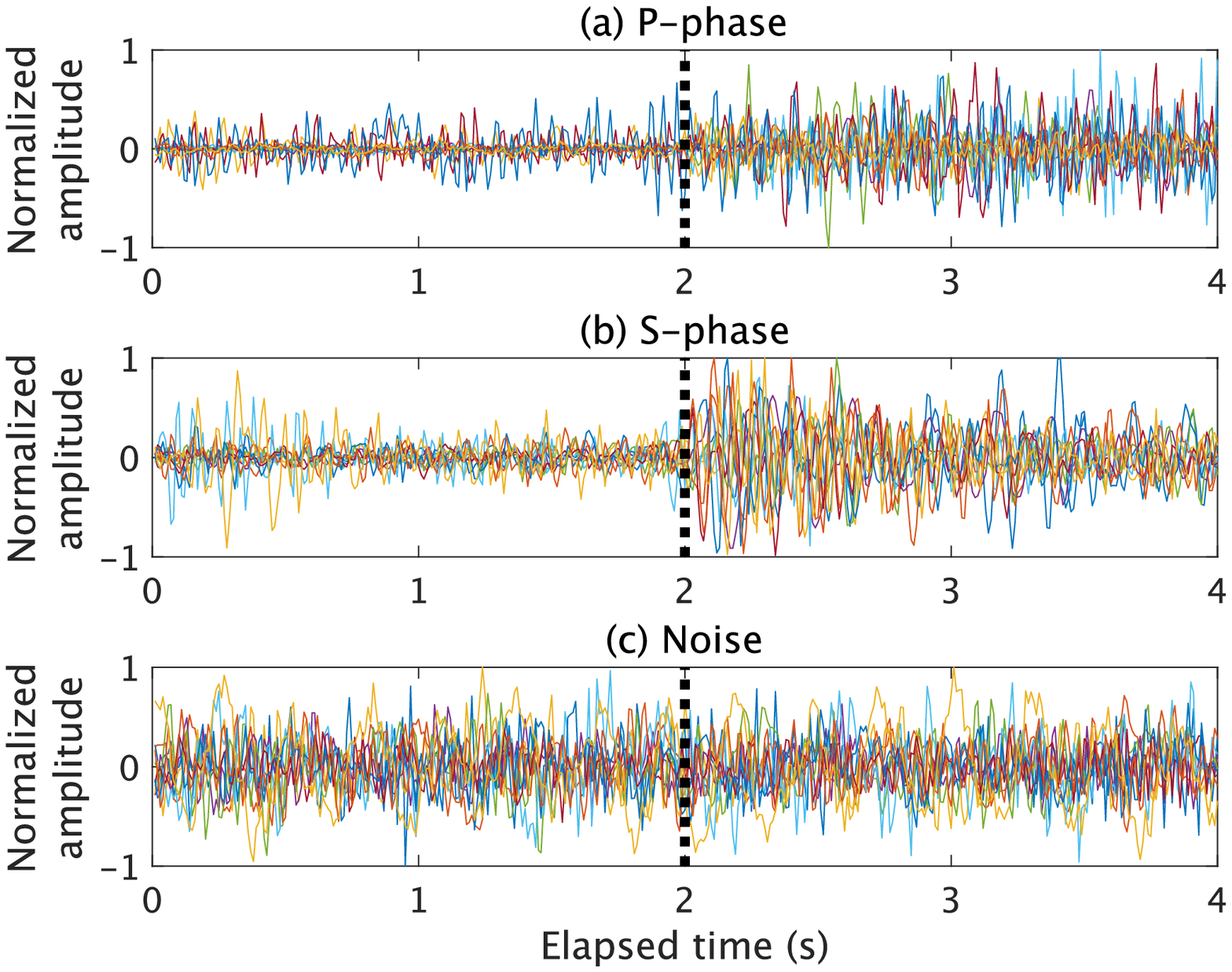}
\includegraphics[scale=0.7, trim=-10mm 0mm 0mm -5mm]{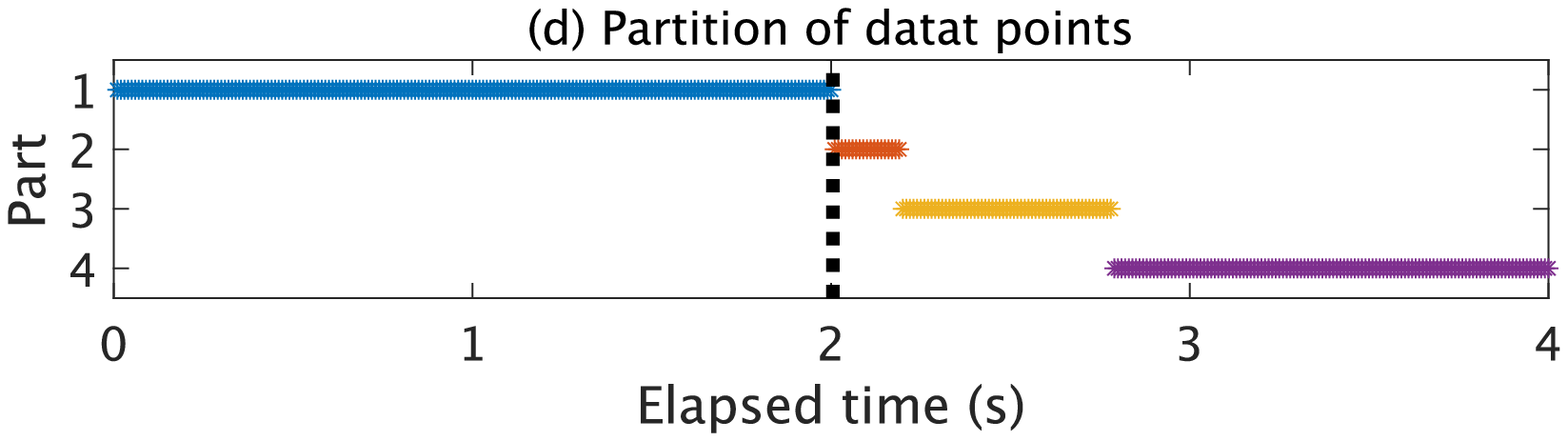}
\caption{Examples of waveforms in the training data \cite{ross2018generalized} and clustering results of data points. Ten waveforms with up-down components were randomly selected and overlayed 
for P-phase (Panel (a)), S-phase (Panel (b)), and noise (Panel (c)), respectively. For the P- and S-phases, the onset time is centred, which is denoted by a black dotted line. The horizontal axis denotes elapsed time (s), whereas the vertical axis denotes normalised amplitude. The normalisation was performed by dividing all records by the maximum of the absolute amplitude in any of the three components (i.e., up-down, north-south, and east-west) \cite{ross2018generalized}. The number of data points is 400 (i.e., a sampled frequency of 100 Hz). 
Panel (d): The optimal partition of data points using multiple clustering. 
We applied multiple clustering for 1000 instances of waveform randomly selected from the training dataset \cite{ross2018generalized}. 
We ran the multiple clustering algorithm for 30 random initial configurations, and 
selected the optimal solution based on MAP (maximum a posteriori ) estimation. The horizontal axis denotes the same data points as in panels (a)-(c), whereas the vertical axis the partition of data points (denoted as \mysingleq{Part}). A single point in the panel corresponds to a data point, which identifies the part to which the data point belongs. Since the majority of adjacent data points belong to the same part, the plotted points appear as a line.}
\label{wave}
\end{figure}

The main idea of our proposed method is to incorporate both global and local waveform information explicitly into a phase-detection model. In the present study, \mysingleq{global representation} refers to 
features related to the entire waveform, whereas the
\mysingleq{local representation} refers to features in a local part of the waveform. To identify local representations, we reviewed the general procedure of seismic-phase detection. In conventional practice, a particular part of the waveform (4–60 s, depending on the method) is extracted from continuous waveform data, which are further normalised (i.e., divided) by the maximum absolute amplitude (e.g., the GPD model) or the standard deviation (e.g., EQT).  This normalisation step plays a crucial role in phase detection by modulating various scales of the waveform. Importantly, its effect is not limited only to the contrast of the waveform before and after the onset, but differences among the three phases also appear in the data points before and after the onset, respectively (Fig.~\ref{wave}). This suggests the possibility that we may better classify the three phases, combining information on the entire part, the before-onset part, and the after-onset part. Thus, the question arises whether these focal parts intrinsically contribute to phase detection. If this is the case, it would be beneficial to make the best use of the information of different focal parts for phase detection. 

We examined this question in a more general framework, in which local representations may be derived in a data-driven manner. To this end, we applied a multiple clustering method \cite{tokuda2017multiple, tokuda2018identification} to waveform data. The multiple clustering method optimally partitions features (data points in our context) into several subsets and clusters instances of the waveform using each subset of features. Hence, if $K$ subsets of features are yielded, $K$ cluster solutions are identified for instances of the waveform, in which $K$ is estimated in a data-driven manner using  the Dirichlet process \cite{gelman2013bayesian,li2019tutorial}.
Importantly, the cluster solutions are assumed to be independent in terms of probabilistic distributions of features. In other words, each cluster solution captures different clustering patterns of instances, which in turn suggest different information on instances (for more details, please refer to Appendix~\ref{multi}). We applied the multiple clustering method to a dataset of 1000 instances of a 4-s waveform (100 Hz, N-S component) including the P-phase, S-phase, and noise. These instances were randomly selected from Southern California Seismic Network (SCSN) data \cite{ross2018generalized}, which have been previously used to train the GPD model. Importantly, in the SCSN data, the data point for the analyst’s selection of P- and S-phases is centred for each instance, as shown in Fig.~\ref{wave}. Hence, it is expected that a common structure of local representations over instances can be derived using the multiple clustering method .

The multiple clustering results suggest that the data points were optimally partitioned into four 
parts (Fig.~\ref{wave}d): The data points in the first half of the segment constitute a single partition, whereas those in the second half constitute three partitions. Furthermore, it was found that the cluster solutions of these four parts are completely different and that they are also rather different from the cluster solution using all data points in terms of the Adjusted Rand Index (ARI) \cite{hubert1985comparing} (see Table~S1 in the Supplementary Material). These results imply that the four partitions of the data points provide different information characterising the instances. This observation motivated us to develop a phase-detection method based on both global (all data points) and local representations (partitions of data points). We expect that a detection method of this kind would be more robust against noise because of the additional information on the waveform. In the present study, instead of the four partitions, we consider local representations characterised by the first and second halves of the data points, which facilitates the implementation of our proposed method in terms of training and testing. It should be noted that these specific local representations coincide with our initial observation of the waveform shown in Fig.~\ref{wave}, as discussed in the first paragraph of this section. 

\section{Model}
Based on the results in Section \ref{intro2}, we developed a phase detection model. In addition to the global representation, we considered local representations that focused on the first and second halves of the waveform. As a basic model, we used the convolutional neural network (CNN) of the GPD model \cite{ross2018generalized}.
For each representation, a different phase detection model was trained. For a new waveform, the phase probability was evaluated as a product of the phase probabilities of each model, assuming that these models should yield consistent phase. 

\subsection{Basic model} \label{basicmodel}
We based our model on the GPD model, which consists of six layers of neural networks: four convolution  layers and two fully connected layers (Table~\ref{tabcnn}). The convolution  layers extract relevant features for phase detection, whereas the fully connected layers classify the waveform based on these extracted features. The input is the 4-s waveform data with a sampling rate of 100 Hz, whereas the output, which is denoted by $P_G(\cdot)$ (the suffix \mysingleq{G} denotes \mysingleq{global}),  is a three-dimensional vector of probabilities for the seismic phases: P-phase (\mysingleq{$p$}), S-phase (\mysingleq{$s$}), and noise (\mysingleq{$n$}). For the P- and S-phases, the phase initiation point was assumed to be at the centre of the window. Hence, the training waveform data should thus be sampled. Because of this training design, for test waveform data, a large phase probability is expected when the (true) initiation point is at the centre, whereas the phase probability decreases as the (true) phase initiation point moves away from the centre. 

\begin{table}
 \caption{ The deep learning model architecture for both global and local representations. They differ only in the convolutional layer’s filter size, which is shown in the fourth row of the table. We follow the abbreviations from \cite{ross2018generalized}: \mysingleq{CBP} denotes convolution, batch normalisation, and pooling (in this order); \mysingleq{FB} denotes fully connected and batch normalisation (in this order); \mysingleq{F} denotes fully connected.}
\vspace{-5mm}
   \begin{flushleft}
    \begin{tabular}{|c|c|c|c|c|c|c|c|}
    \hline
     Layer & 1 & 2& 3& 4& 5& 6& 7\\
     \hline
    Stage & CBP & CBP & CBP& CBP& FB& FB& F \\
    \hline
    Number of channels & 32& 64& 128& 256& 200& 200& 3 \\
    \hline
    Filter size (Global/local)  &21/10 & 15/7 & 11/5 & 9/4 & - & - & - \\
    \hline
    \end{tabular}
    \end{flushleft}
    \label{tabcnn}
\end{table}

\subsection{Proposed model}
Inspired by the multiple clustering results in Section \ref{intro2}, we considered separately developing phase detection models for both global and local representations. For global representation, we adopted the GPD model, in which the waveform of all data points was taken as the  input. Conversely, for  local representations, we considered a model in which the waveform of the first  or  second half of the data points was taken as the input (Fig.~\ref{architecture}). For these local models, the architecture of the neural network was basically identical to the global one (Table~\ref{tabcnn}), but the filter size was reduced by  half because the window size was  narrower. For these local models, we denoted the phase probabilities for the first and second halves of the data points as $P_{L1}(\cdot)$ and $P_{L2}(\cdot)$ (the suffix \mysingleq{L} denotes \mysingleq{local}), respectively. Finally, we defined the final output $P(x)$ as the product of the three probabilities yielded by the global model and the two local models:  
\begin{eqnarray}
P(x) \overset{def}{=} P_{G}(x)^{w_G} \times P_{L1}(x)^{w_{L1}} \times P_{L2}(x)^{w_{L2}},
\label{probdef}
\end{eqnarray}
where $x \in\{\mysingleq{p}, \mysingleq{s}, \mysingleq{n} \} $; $w_G, w_{L1}, w_{L2} \in \{0, 1\}$. The binary parameters $w_G, w_{L1}$ and $w_{L2}$ allowed us to widen the scope of the model representations.
The proposed model, hereafter referred to as the \mysingleq{GL model}, can be expressed by setting binary parameters as follows:
\begin{itemize}
\item GL model: $(w_G, w_{L1}, w_{L2})=(1, 1, 1)$.
\end{itemize}
Similarly, the three  components of the GL model
can be expressed as the following sub-models:
\begin{itemize}
\item G model: $(w_G, w_{L1}, w_{L2})=(1, 0, 0)$
\item L1 model: $(w_G, w_{L1}, w_{L2})=(0, 1, 0)$
\item L2 model: $(w_G, w_{L1}, w_{L2})=(0, 0, 1)$,
\end{itemize}
where model G is the global model, L1 is the local model for the first half of the data points, and L2 is the local model for the second half of the data points. The G model is identical to the GPD model. These sub-models are separately trained using the training data (inputs 1-3 and outputs 1-3 in Fig.~\ref{architecture}), whereas the final output in Eq.(\ref{probdef}) is evaluated only for the test data. 

Because of the formulation of the GL model,  the summation of probabilities is not always one,  that is, $P(\mysingleq{p}) + P(\mysingleq{s}) + P(\mysingleq{n}) \neq 1$.
One may consider normalising the probabilities, but in the present study, we retained this formulation as it is because an interpretation of probability $P(x)$ is straightforward. It should be noted that the normalisation does not change the large and small relationships among the probabilities of $x \in\{\mysingleq{p}, \mysingleq{s}, \mysingleq{n} \}$ (because $a \leq b \Leftrightarrow a/(a+b) \leq b/(a+b)$ for $a>0$ and $b>0$). Hence, the classification results (Sections \ref{class1} and \ref{class2}) based on the maximum likelihood principle remained unchanged regardless of whether normalisation was performed.

\subsection{Model training} \label{training}
We trained the G, L1, and L2 models using the waveform dataset of the Southern California Seismic Network (SCSN),  which was used to train the GPD model in \cite{ross2018generalized}. Based on these sub-models, we constructed the GL model, as shown in Eq.(\ref{probdef}). 
The training dataset consists of 4.5 million instances, which are equally distributed for the P-phase, S-phase, and noise. All the waveforms were sampled at 100 Hz for 4 s. For the P-phase and S-phases, the waveform was centred on the analyst’s selection. We randomly split the data into three subsets: the training data (75 \%), validation data (15 \%), and test data (10 \%). The validation data were used for early stopping, the conventional mechanism used in deep learning to prevent overfitting \cite{goodfellow2016deep, geron2019hands}. 
To train the model, we followed the same settings as in \cite{ross2018generalized}: 480 batch size, batch normalisation, early stopping with five patience epochs, the cross-entropy loss function, and the Adam optimisation algorithm. It was found that only a few epochs were sufficient for the training to converge (Fig.~S1 in the Supplementary Material). Using four NVIDIA GPUs in our computational environment, the training took less than 30 min. 
 
\section{Application}
Based on the trained models described in the previous section, we applied the proposed method to various types of waveform data. 
First, we report the classification results of the test data using the GL, G, L1, and L2 models. Subsequently, we examined the classification performance when the test data were noise-contaminated. Third, we applied the proposed method to continuous seismic waveform data, taking the example of the 2016 Bombay Beach swarm. Finally, we applied the method to continuous seismic waveform data in Japan, which includes LFEs. LFE is a specific type of earthquake that dominates lower frequency component (2-8 Hz). In the latter two cases, the performance was compared with that of the GPD model, PhaseNet, and EQT.

\subsection{Classification of test data}\label{class1}
\begin{figure}
\includegraphics[scale=0.40, trim = 25mm 0mm 0mm 0mm]{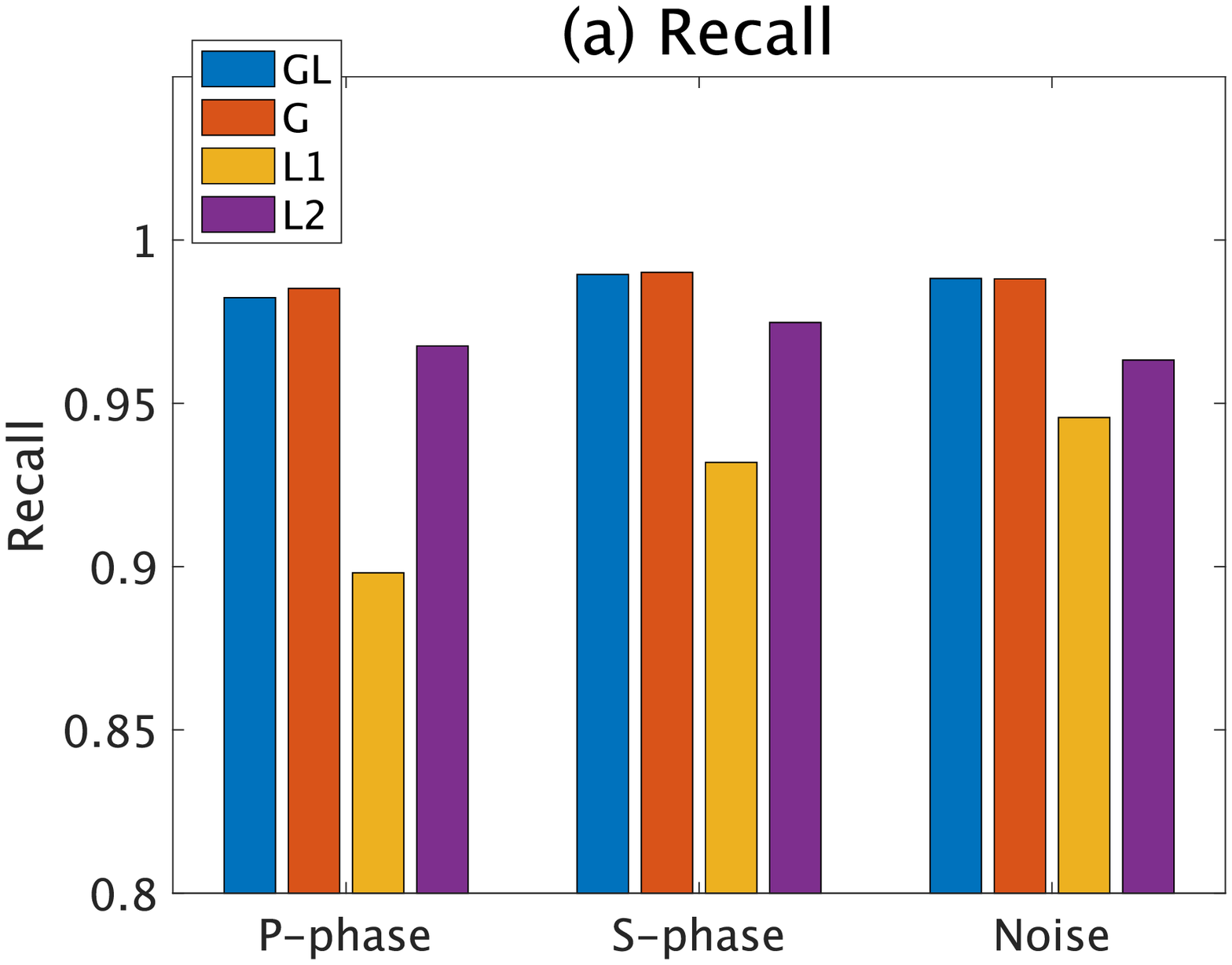}
\includegraphics[scale=0.40, trim = 0mm 0mm 0mm 0mm ]{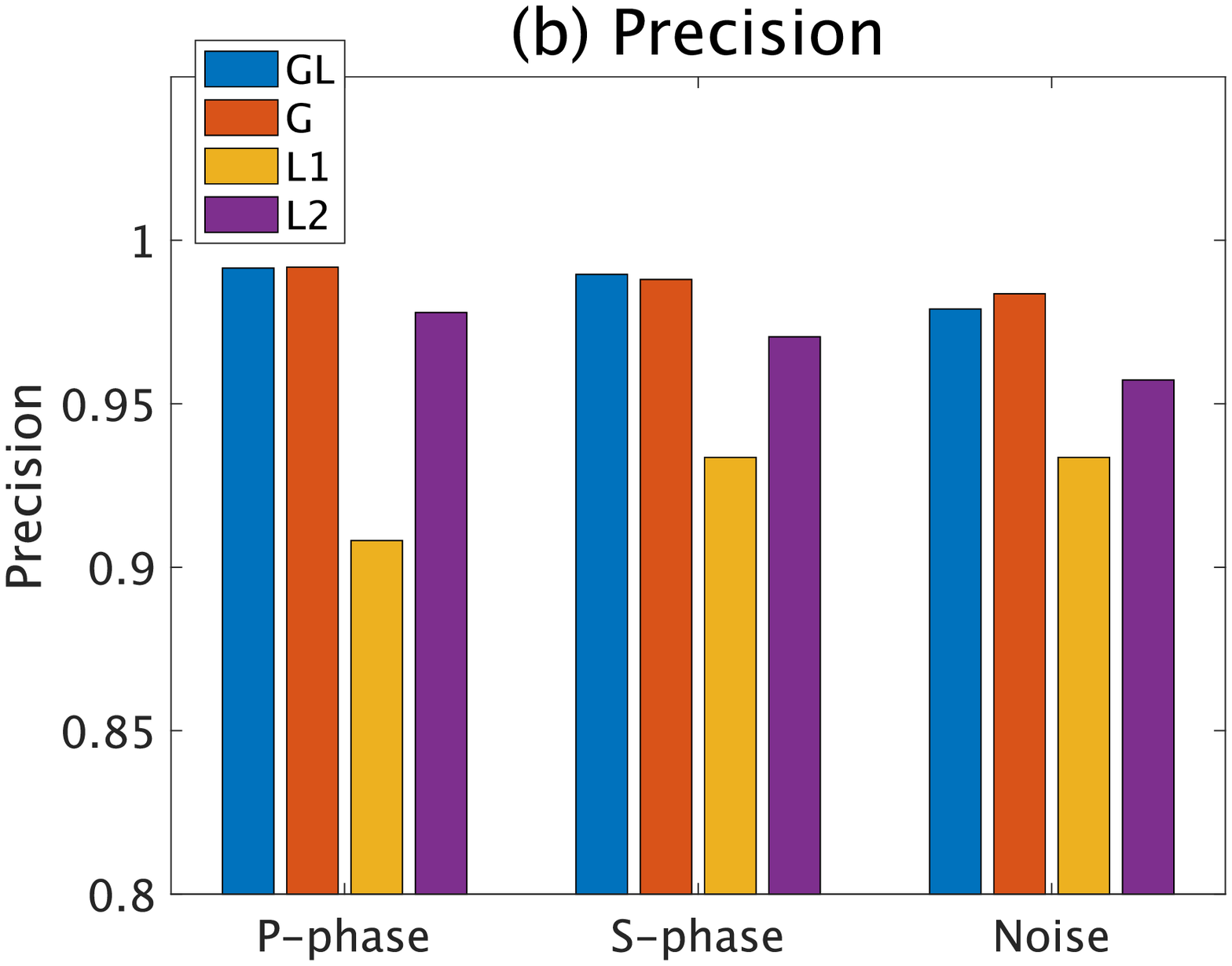}
\caption{Performance of classification for the test data. Panel (a): Recall. Panel (b): Precision. Both panels summarise the  P-phase, S-phase, and noise detection performance. The color bars denote the performance for the GL model (\mysingleq{GL}), G model (\mysingleq{G}), L1 model (\mysingleq{L1}), and L2 model (\mysingleq{L2}), respectively. For more classification results details, please refer to Table~S2 in the Supplementary Material.}
\label{recall}
\end{figure}

We report on the classification results of the three phases for the test dataset in Section~\ref{training}. The classification is based on the maximum likelihood principle by which the phase that provides the maximum probability is assigned. We applied the GL, G, L1, and L2 models to the test data, which were evaluated in terms of recall and precision (Fig.~\ref{recall}). Here, we adapted the recall and precision metrics \cite{buckland1994relationship} for multiclass classification, that is, \mysingleq{p}, \mysingleq{s}, \mysingleq{n}: We extended the definition of completeness of retrieval (recall) and purity of retrieval (precision) by focusing on a particular class (for more details, please refer to Appendix~\ref{recalldef}). 
It was found that the GL and G models performed equally well, with the recall and precision of both being $>$0.98. The performance of the L2 model was also good, with a recall and precision of $>$0.95, whereas the recall and precision of the L1 model deteriorated slightly, scoring between 0.90 and 0.95. These results suggest that the global (i.e., the G model), and the local model (i.e., the L1 and L2 models) offer considerable information for phase detection, as shown in Fig.~\ref{wave}. 

\subsection{Classification of noise-contaminated data}\label{class2}
Now, the question arises concerning whether there is any difference between the GL and G models in terms of phase detection capability. 
Because the GL model contains additional information in the first and second halves of the waveform, it was hypothesised that the GL model may provide more robust phase detection than the G model. To test this hypothesis, we performed a simulation study in which noise was added to the test data. Here, we denote \mysingleq{noise} by a noise-phase waveform. 
Denoting $\gamma$ as the noise proportion ($0 \leq \gamma \leq 1$), we considered transforming the waveform by: 
\begin{eqnarray}
\bb{x} \leftarrow (1-\gamma) \times \bb{x} + \gamma \times \bb{n},
\label{noiseratio}
\end{eqnarray}
where $\bb{x}$ is the waveform data with arbitrary phase and $\bb{n}$ noise. This additive noise model is a natural and simplified approximation, but it is tractable and good enough for the test.
For this simulation study, we generated the following datasets: First, 1000 waveforms were randomly selected from the test data. Second, each instance was transformed into Eq.(\ref{noiseratio}) using instance $\bb{n}$, which was randomly and independently sampled from the test data that were not selected for the aforementioned 1000 instances. We manipulated the value of noise proportion $\gamma$ from zero to one in increments of 0.25. Moreover, to evaluate the effect of the locus of noise contamination, we considered the following three cases: all data points, the first half of the data points (DP1), and the second half of the data points (DP2). In the case of contamination in DP1, 
the data transformation in Eq.(\ref{noiseratio}) is performed only for DP1, whereas DP2 remain unchanged. Similarly, in the case of DP2, contamination is limited to DP2, whereas DP1 remain unchanged. Hereby, we generated 100 datasets for each value of noise proportion $\gamma$ and the locus of contamination. For these datasets, we applied the GL, G, L1, and L2 models to evaluate their performance in terms of accuracy, which summarises overall classification performance (please refer to Appendix~\ref{recalldef} for the definition of accuracy). 

\subsubsection*{Results}
\begin{figure}
\includegraphics[scale=0.4, trim = 25mm 0mm 0mm 0mm]{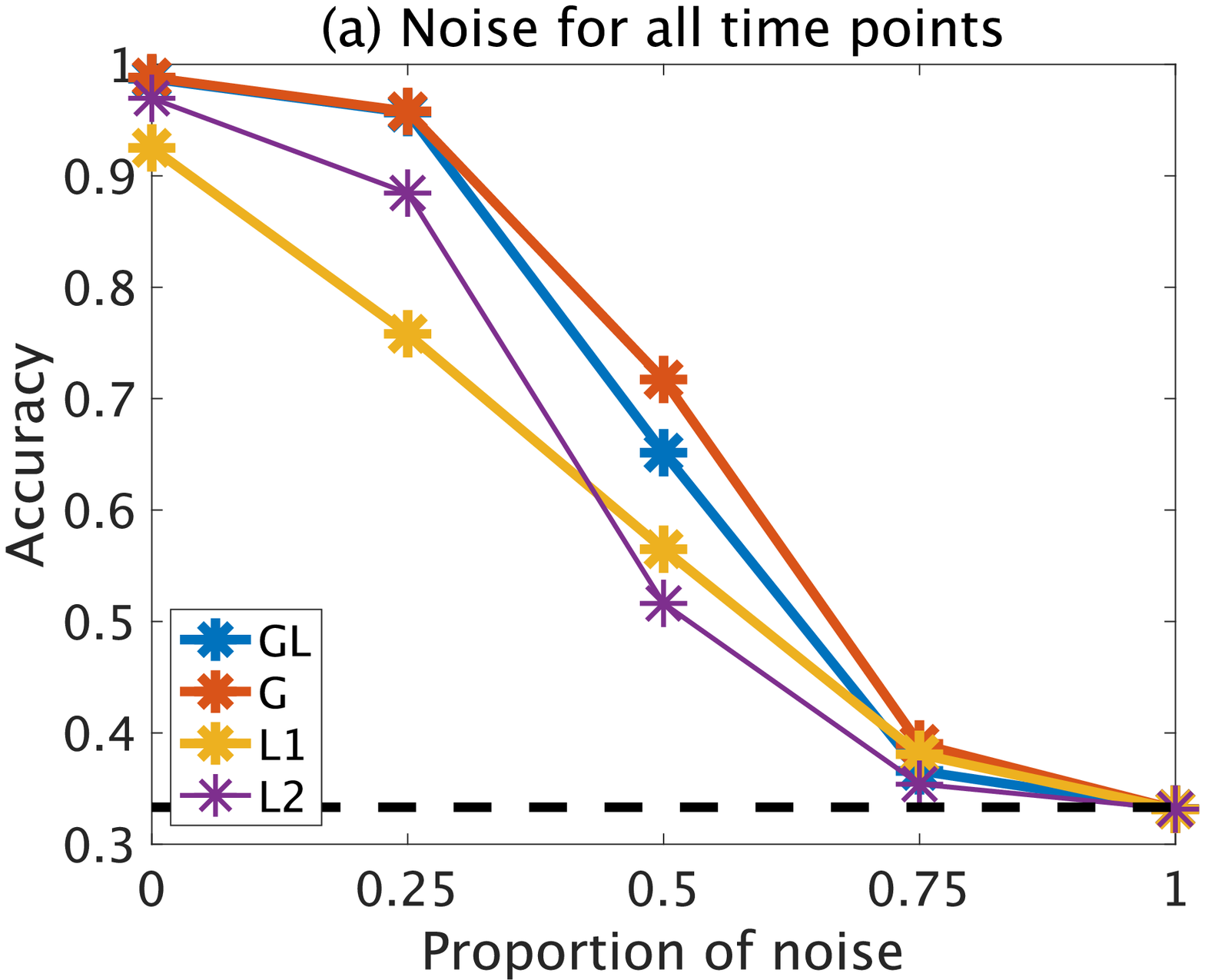}
\includegraphics[scale=0.4, trim = 0mm 0mm 0mm 0mm]{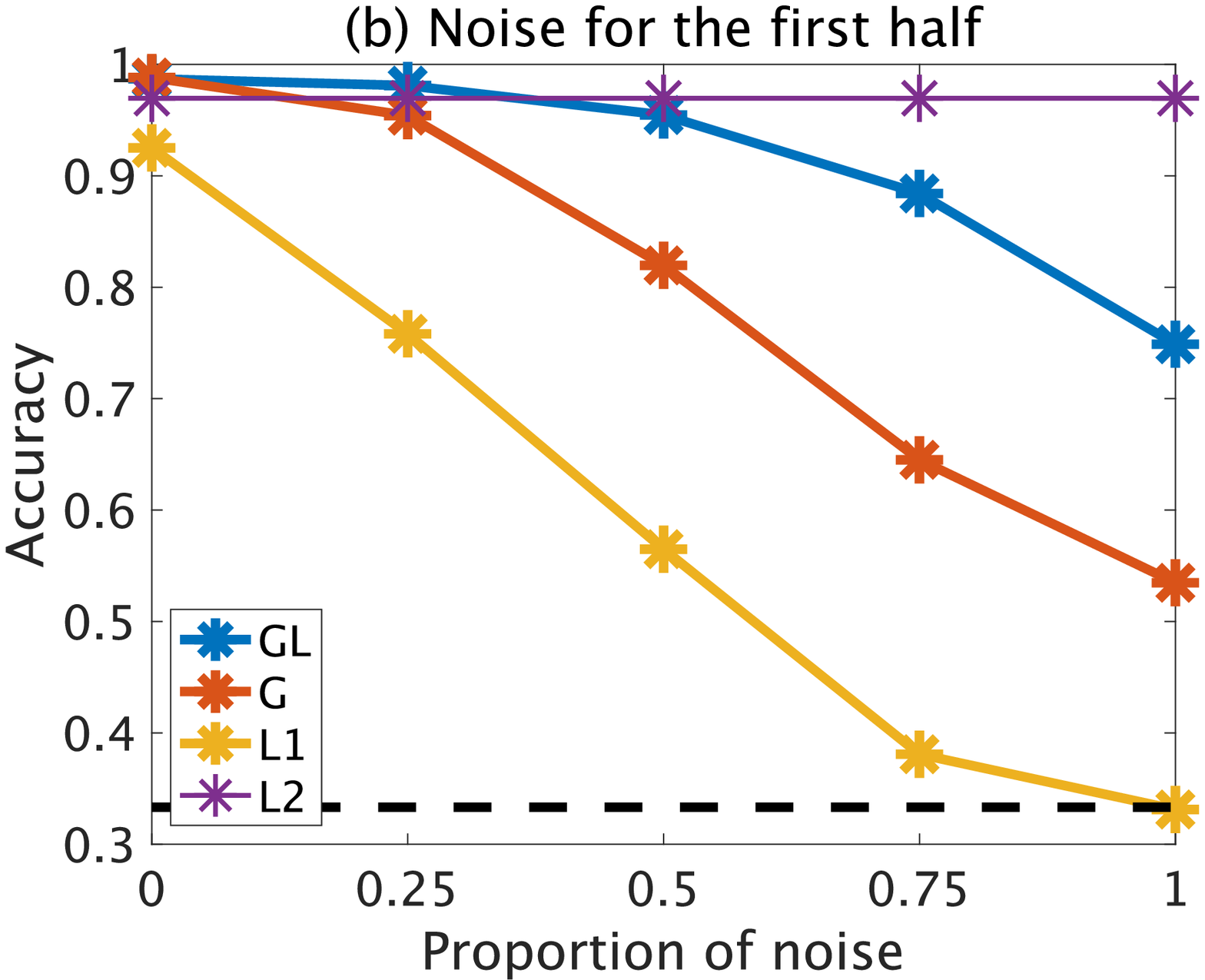}
\includegraphics[scale=0.4, trim = 25mm 0mm 0mm -10mm]{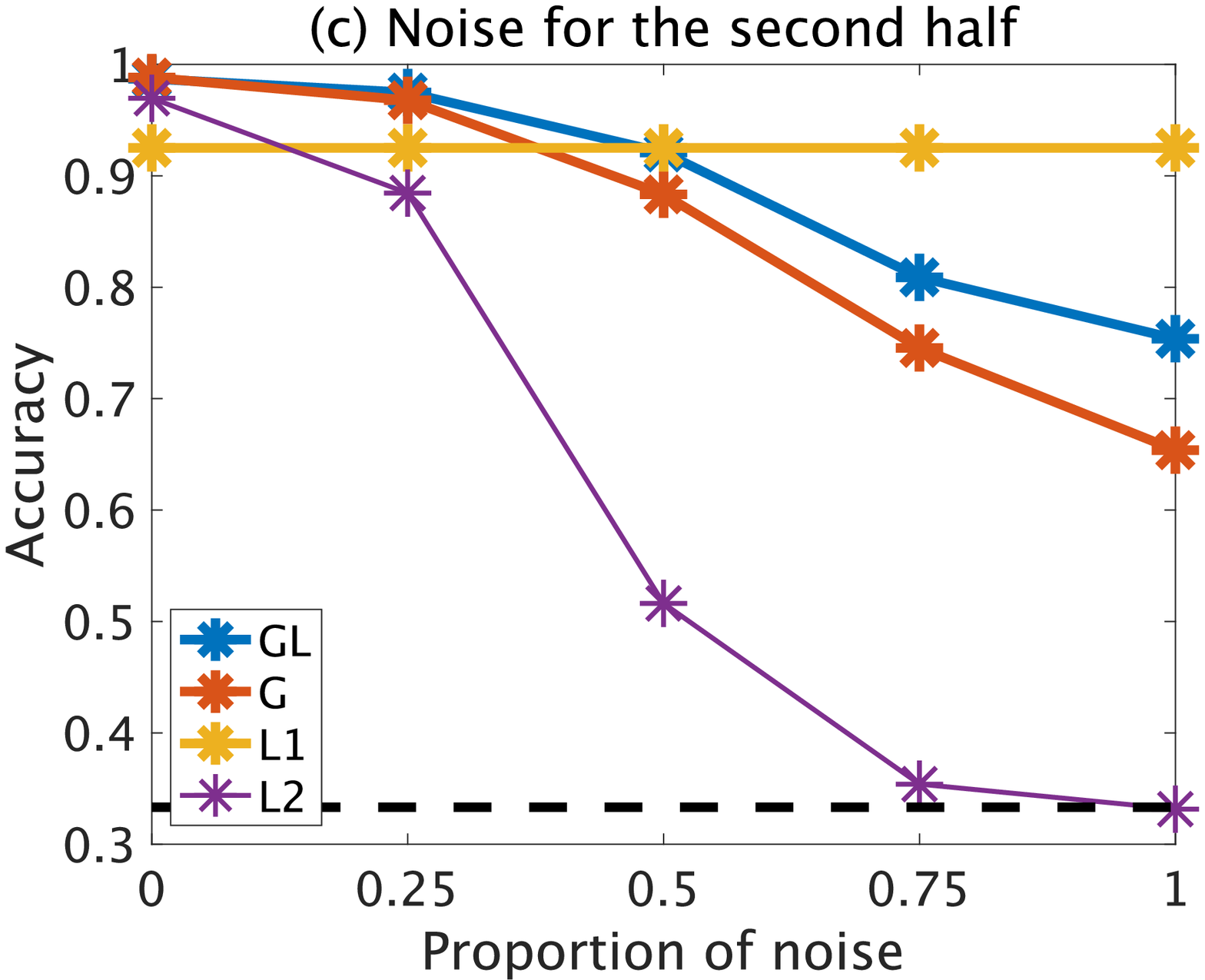}
\caption{Classification performance for noise-contaminated test data. Panel (a): The noise waveform is added to all data points. Panel (b): The noise waveform is added to the first half of the data points. 
Panel (c): It is added to the second half of the data points. The horizontal axis denotes noise proportion, whereas the vertical axis denotes mean accuracy for 100 datasets of 1000 randomly selected instances. The black dashed line denotes the expected accuracy at the chance  level (i.e., 1/3). The color lines denote performance for the GL model (\mysingleq{GL}), G model (\mysingleq{G}), L1 model (\mysingleq{L1}), and L2 model (\mysingleq{L2}), respectively.
For the mean values and standard deviations, please refer to Table~S3 in the Supplementary Material.
}
\label{performancecontami}
\end{figure}

When noise was added to all data points, the accuracy largely decreased for all models as the noise proportion $\gamma$ increased (Fig.~\ref{performancecontami}a). In this case, the performance of the GL model (blue in the figure) was slightly worse than that of the G model (red in the figure) for a noise proportion of 0.5 with accuracy 0.72 and 0.65, respectively. Similarly, when noise was added only for the first half of the data points, the accuracy tended to decrease for the GL, G, and L1 models as the noise proportion $\gamma$ increased.   
However, the accuracy of the L2 model remained high irrespective of the noise proportion $\gamma$ because noise no longer influences the L2 model (Fig.~\ref{performancecontami}b). Importantly, the performance of the GL model was significantly better than that of the G model: For a noise proportion of 0.75 the accuracy of the GL model and G model was 0.88 and 0.65, respectively. 
A similar observation was made when noise was added to the second half of the data (Fig.~\ref{performancecontami}c).

\begin{figure}
\includegraphics[scale=0.7]{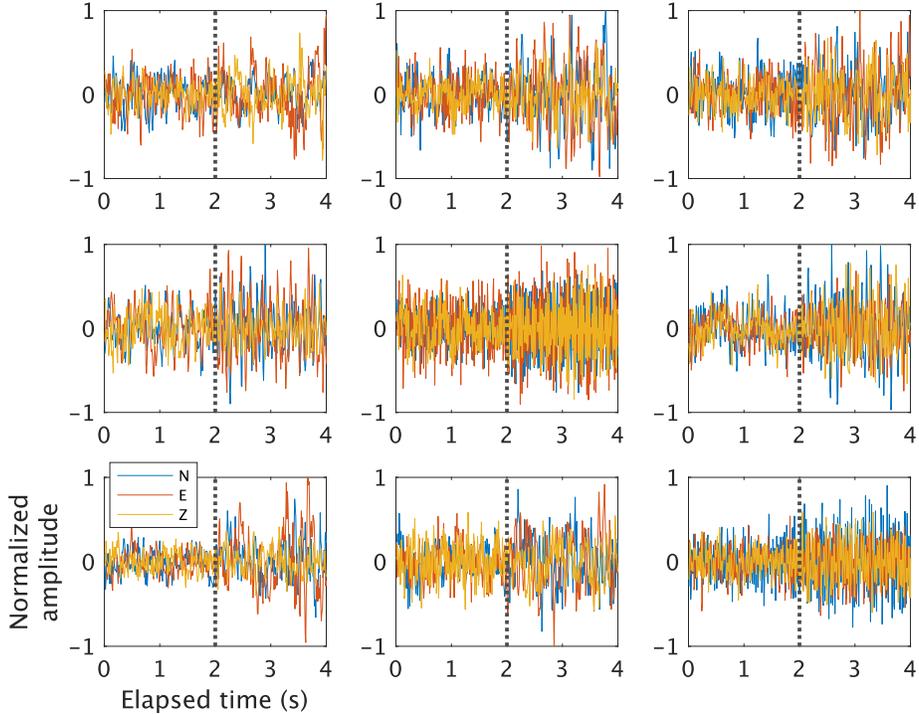}
\caption{Examples of waveforms that the GL model correctly identified as noise that the G model misidentified. These waveforms are contaminated by noise with a noise proportion of 0.5 in Eq.(\ref{noiseratio}). The horizontal axis denotes elapsed time (4 s), whereas the vertical axis denotes normalised amplitude (we did not re-normalise the waveform after the transformation in Eq.(\ref{noiseratio})). The dotted line denotes the middle point of the time window. For the legend, \mysingleq{N} denotes N-S component; \mysingleq{E} E-W component; \mysingleq{Z} up-down component.
These examples were randomly selected from instances used in Section \ref{class2}. 
}
\label{failedcases}
\end{figure}

Next, we examined waveform instances that the G model misclassified and the GL model correctly classified. As a typical case, we focused on the setting in which noise was added to the first half of the data points with a noise proportion of 0.5. In this setting, such instances most frequently occurred in the noise phase (the average number of such instances was 31.8, 39.4, and 69.1 out of 1000 instances for P-phase, S-phase, and noise, respectively). 
We visually investigated the waveforms of these instances (Fig.~\ref{failedcases}). It was found that  there were differences in amplitude between the first and second halves of the data points in which the second half had a larger amplitude. Possibly because of this apparent contrast in amplitude
between these two parts, the G model misclassified the noise as either the P- or S-phase. Conversely, the GL model worked well for these instances because it had information on the L1 and L2 models, which correctly classified the noise phase. 

\subsection{2016 Bombay Beach swarm} \label{sectionbombay}
Subsequently, we applied the proposed method (GL model) to the 2016 Bombay Beach swarm that occurred in California, USA  \cite{ross2018generalized, mcbride2020earthquakeadvisory}. 
The objective was to evaluate the performance of the proposed method for continuous seismic waveform data, which was further compared with the performances of the GPD model and two other state-of-the-art methods based on deep neural networks: PhaseNet \cite{zhu2019phasenet} and 
EQT \cite{mousavi2020earthquake}. For these methods, we used publicly available open-source models from \cite{woollam2022seisbench}, by which the GPD model was trained using the original training data from Southern California, whereas PhaseNet and EQT were trained using the SCEDC (Southern California Earthquake Data Center) dataset. Thus, the geometric region for the training data coincided with the location of the Bombay Beach swarm for all these methods, including the proposed method. Hence, it was expected that these methods would be appropriate for detecting the 2016 Bombay Beach swarm phases.

We set the detection thresholds for both the P- and S-phases as follows: 
For the GL model, we set the phase probability to 0.5. This criterion implies that all three sub-probabilities on the right-hand side of Eq.(\ref{probdef}) should be greater than 0.5. If these probabilities are identical, each probability becomes 0.79. For the remainder of the methods, we adopted thresholds from their original papers: 0.98 for the GPD model \cite{ross2018generalized}, 0.5 for PhaseNet \cite{zhu2019phasenet}, and 0.3 for EQT  \cite{mousavi2020earthquake}. It is noted that in their original papers, no clear reasons were given for these thresholds other than empirical ones. 
The window size for phase detection followed the original setting: 4 s for the GL and GPD models, 30 s for PhaseNet, and 60 s for EQT. Furthermore, we set the shift width (stride) to 0.1 s for all methods.

We applied these detection methods to three-component waveform data recorded at a frequency of 100 Hz at the BOM station in the Southern California Seismic Network \cite{https://doi.org/10.7914/sn/ci, SCEDC}.
From the waveform record, we extracted 6-h waveforms before and after the onset of the swarm. However, to avoid uncertainty in the swarm’s onset time, we removed the waveform data from 1 h before and after the onset, which produced a 5-h waveform for each time segment.  
To alleviate the possible effect of earthquakes that occurred before the onset of the swarm, we discarded the detection results 1 min after the earthquakes in the SCEDC catalogue. In this catalogue, 12 earthquakes were recorded during this period, 11 of which had magnitudes of less than 1.5, with the hypocentre more than 65 km away from the BOM station. One earthquake had a magnitude of 6.0, but it took place 10440 km away from the station. Because of these data selection procedures, in the following analysis, we assumed that there was no effect attributable to the earthquakes/swarm before the onset.

\subsubsection*{Results}
First, a visual inspection of the phase detection suggests substantial differences in the P- and S-phase detections before and after the onset of the swarm 
(Figs.~\ref{bombayall}, \ref{bombayZoom1}, and \ref{bombayall2}). For all methods, more P- and S-phases were detected after the onset than before it. However, among these methods, there were considerable differences in the phase-detection efficacy. Such differences after the onset are clearly shown in Fig.~\ref{bombayZoom1}. 
Furthermore, we quantified the number of phase detections before and after the onset (Table~\ref{tabswarm}).
With regard to the P-phase, the GL model yielded two  detections before onset, whereas 923 detections occurred after onset. In contrast, the GPD model yielded 18 detections before onset, whereas 1336 detections occurred after onset. 
PhaseNet yielded no detections before onset, whereas 67 detections occurred after onset. Finally, EQT yielded seven detections  before the onset, whereas 353 detections occurred after the onset. As the number of detections does not necessarily denote the number of events, it is not straightforward to evaluate these results. Furthermore, among these methods, there were differences in how a detection probability was assigned to non-peak times.
Nonetheless, the ratio of the number of detections before and after the onset (column \mysingleq{Ratio} in Table~\ref{tabswarm}) suggests that the GL model and PhaseNet performed very well, with a ratio of 461 and $\infty$, respectively. However, notably, PhaseNet tended to yield somewhat conservative detection, which did not detect the P-phase in the period of 2 h after the onset (the 8th column in Table~\ref{tabswarm}). Regarding the S-phase, the GL model worked as intended without any detection before the onset of the swarm. 

\begin{landscape}
\begin{figure}
\includegraphics[scale=0.30, trim=100mm 50mm 0mm 50mm]{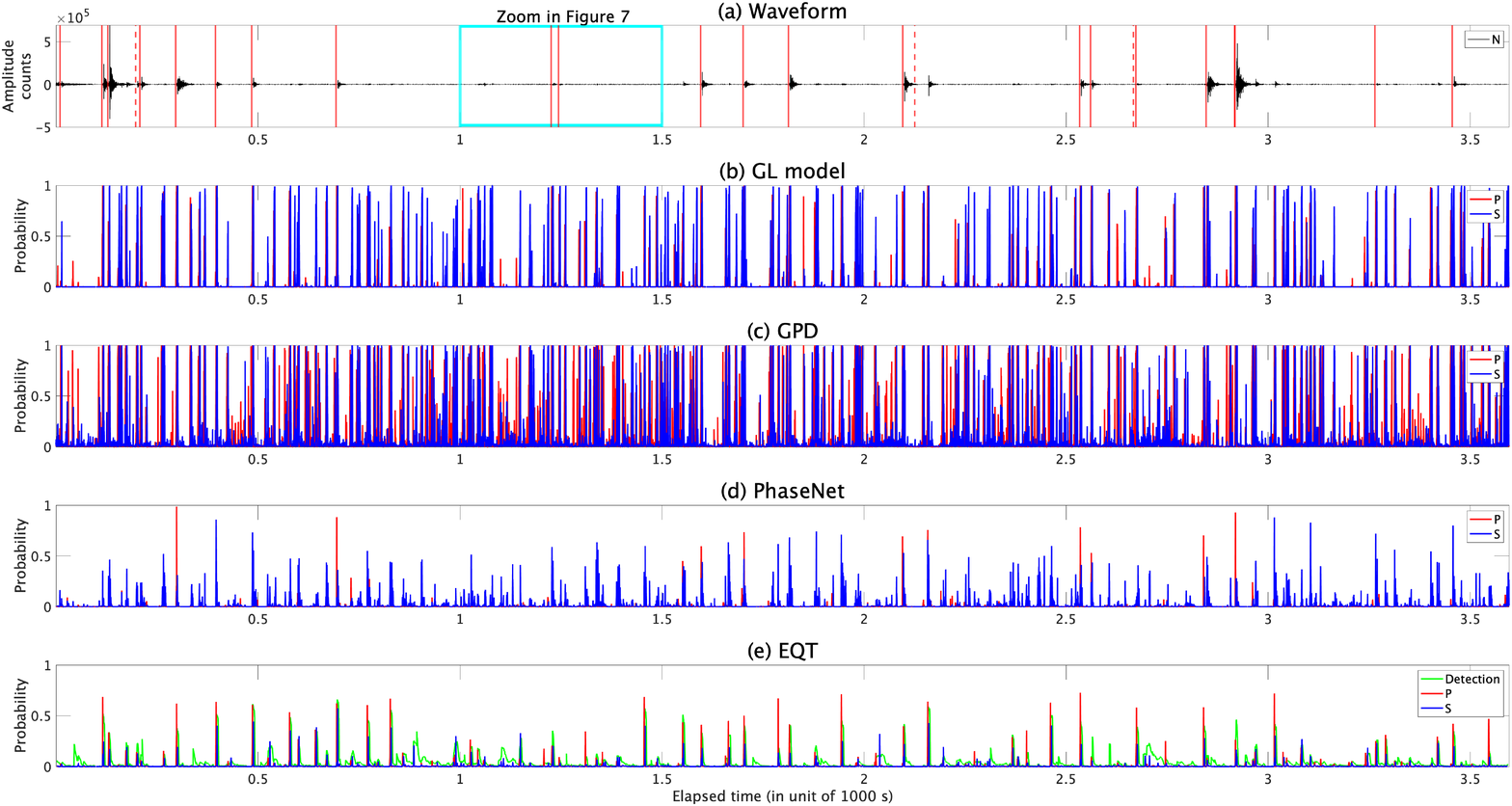}
\vspace{3mm}
\caption{Waveform and phase detections \textit{after} the onset of 2016 Bombay Beach swarm, California, USA. Panel (a): Waveform at the BOM station with the HHN  channel. The horizontal axis denotes the 1-h time course starting at 8:00. Pacific daylight time (PDT), 5 h after the onset of the swarm, whereas the vertical axis denotes amplitude counts. 
The vertical red line denotes the origin time for the earthquakes that occurred less than 10 km from the BOM station in the SCEDC catalogue \cite{SCEDC}, whereas the red dashed lines denote those further than 10 km. The region enclosed by the cyan lines is zoomed in Fig.~\ref{bombayZoom1}. Panels (b)-(e): Phase detection results for GL model, the GPD model, PhaseNet, and EQT, respectively. Phase detection is based on three components of waveform with the HHZ , HHN, and HHE  channels. We pre-processed the waveform data by detrending and high-pass filtering above 2 Hz as in \cite{ross2018generalized}. 
For these panels, the vertical axis denotes the P-wave (in red) and S-wave (i.e., Secondary wave, in blue) probabilities. For EQT, the detection probability is also plotted in green. Except for the GL model, we used the open-source models in the Seisbench toolbox \cite{woollam2022seisbench}.}
\label{bombayall}
\end{figure}

\begin{figure}
\includegraphics[scale=0.30, trim=15mm 50mm 0mm 50mm]{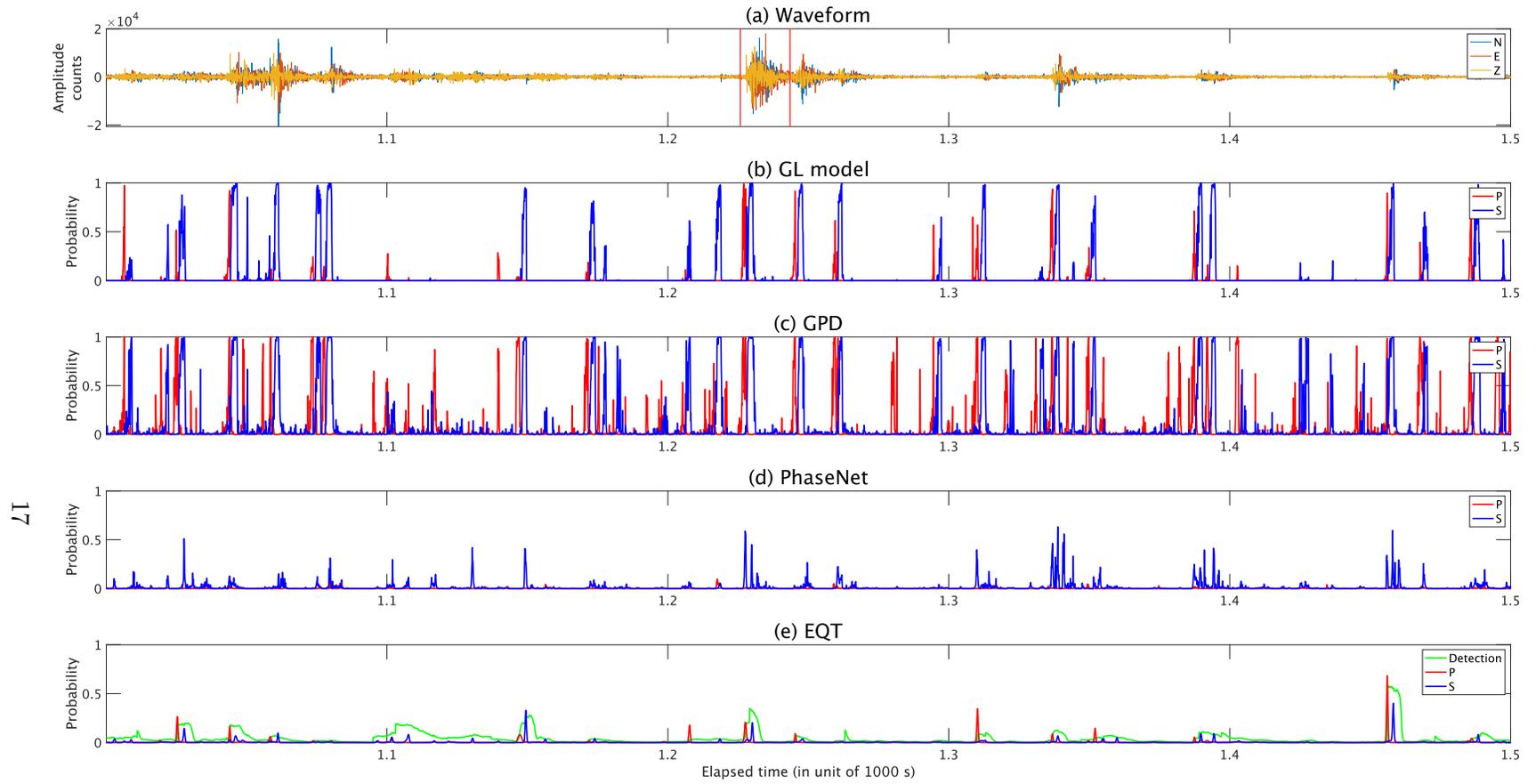}
\vspace{15mm}
\caption{Waveform and phase detections \textit{after} the onset of the 2016 Bombay Beach swarm, which zoomed the time duration from 1000 to 1500 s in Fig.~\ref{bombayall}, which is enclosed by the cyan lines in the figure. For the legend in Panel (a), \mysingleq{N} denotes the N-S component (HHN channel); \mysingleq{E} E-W component (HHE channel); \mysingleq{Z} up-down component (HHZ channel).}
\label{bombayZoom1}
\end{figure}

\begin{figure}
\includegraphics[scale=0.30, trim=100mm 50mm 0mm 50mm]{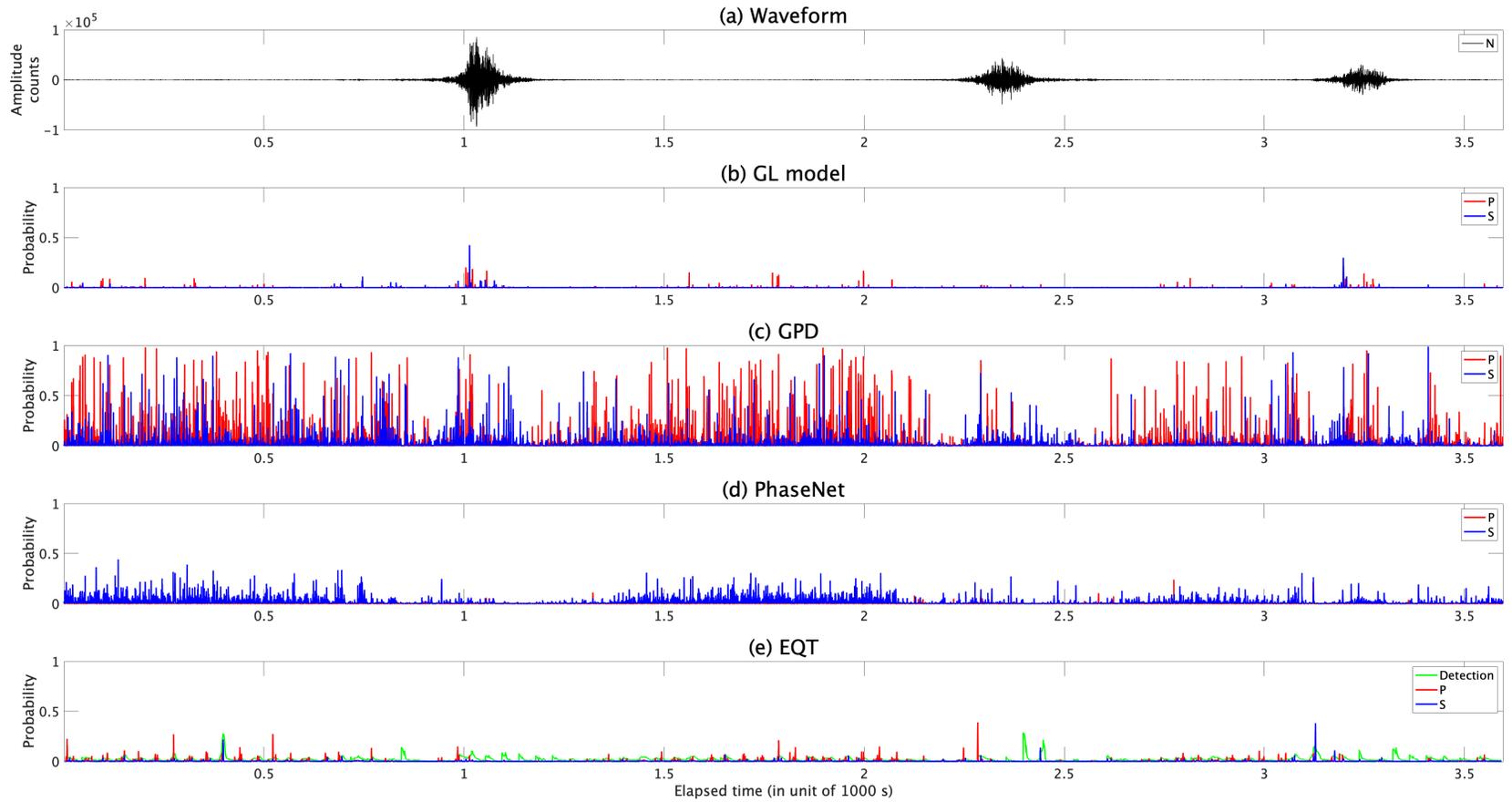}
\vspace{5mm}
\caption{Waveform and phase detection \textit{before} the onset of 2016 Bombay Beach swarm. This figure is similar to Fig.~\ref{bombayall}, but  the time coverage starts 5 h before the onset of the swarm, i.e., at 23:00  on September 25 (PDT). 
}
\label{bombayall2}
\end{figure}
\end{landscape}

\begin{table}
 \caption{Summary of phase detections for the 2016 Bombay Beach swarm. The numbers of phase detections per hour are summarised for the time coverages of 2 and 6 h before and after the onset of the swarm, respectively. 
 In the table, the period before the onset is shown as negative, whereas the period after the onset is positive. 
 The numbers of detections for the P- and S-phases are separately presented. 
 The summation (column \mysingleq{Sum}) is taken over each time segment, i.e., 
 before and after the onset (\mysingleq{before-segment} and \mysingleq{after-segment}).
 The ratio (column \mysingleq{Ratio}) is evaluated as the number of detections in the after-segment over the number of detections in the before-segment. }
 \vspace{-5mm}
 \footnotesize
    \begin{flushleft}
    \begin{tabular}{|c|cccccc|cccccc|c|}
    \hline
     Method & \multicolumn{6}{c|}{\textbf{Before-segment}} & \multicolumn{6}{c|}{\textbf{After-segment}}
     & \textbf{After/Before} \\
     \hline
      Period (h) & -6 & -5& -4& -3& -2& Sum &2&  3& 4& 5& 6 & Sum & Ratio\\
     \hline
     \multicolumn{14}{|c|}{P-phase detection}\\
     \hline
     GL &  0  &  0 & 1 &  0 & 1 & \textbf{2} & 45 &  88 &  210 &  356 &  224 & \textbf{923} & \textbf{461} \\
     GPD & 11 &  1 & 1 & 3 &  2 & \textbf{18} &  51 & 66  &  447 &  430 &  342 & \textbf{1336} & \textbf{74} \\
     PhaseNet & 0 & 0 & 0 & 0  &   0  &  \textbf{0}  & 0 & 9  &  5  &  32  &  21 & \textbf{67} & $\boldsymbol{\infty}$\\
     EQT &  4  &   3    & 0 & 0 & 0 & \textbf{7} & 22 &   36    & 42 &  147 &  106 & \textbf{353} & \textbf{50}\\
      \hhline{|==============|}
      \multicolumn{14}{|c|}{S-phase detection}\\
     \hline
      GL  & 0 & 0   & 0 & 0 & 0 & \textbf{0} & 168 &  594 & 1583 & 1772 & 1550 & \textbf{5667} & $\boldsymbol{\infty}$\\
     GPD & 0 & 1 &  0  & 3  & 0 & \textbf{4} &110  & 400 & 914 & 1114 & 943 & \textbf{3481} & \textbf{870}\\
     PhaseNet & 0 &  0 &  0 & 1 & 0 & \textbf{1} & 8  & 22  &  59 & 59 &  66 & \textbf{214} & \textbf{214}\\
     EQT &  1 & 4 & 0 & 0 & 0 & \textbf{5} &9 & 14 & 20 & 4 & 37 & \textbf{120} & \textbf{24}\\
     \hline
    \end{tabular}
    \end{flushleft}
    \label{tabswarm}
\end{table}

\begin{figure}
\includegraphics[scale=0.18, trim=120mm 0mm 0mm 0mm]{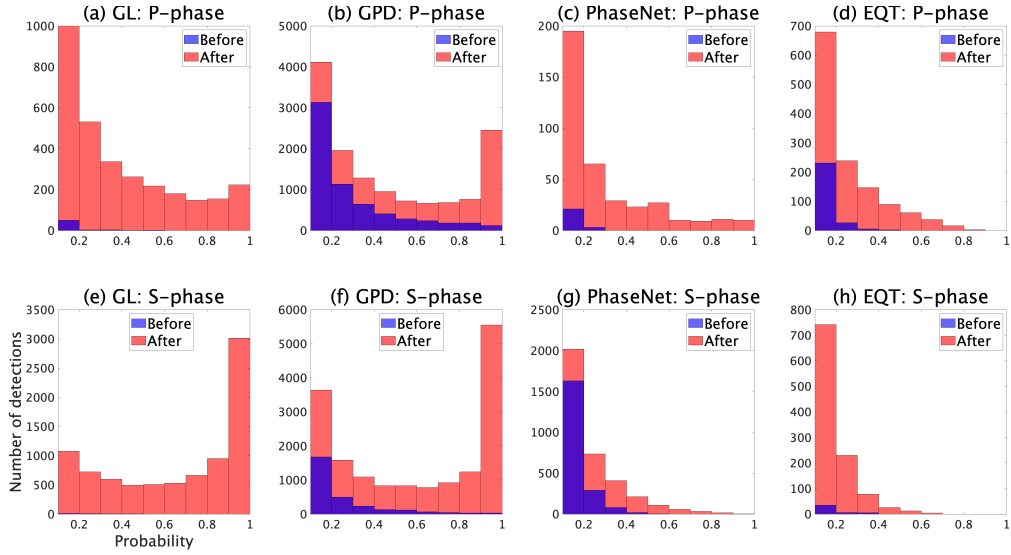}
\caption{The probabilities for P- and S-phase detections before and after the onset of the Bombay Beach swarm. Each panel denotes a histogram for the number of phase probabilities in the 5-h period before the onset (blue bar) and in the 5-h period after the onset (red bar). The horizontal axis denotes the probability for phase detection, and the vertical axis denotes the number of detections. The first and second rows are for the P-phase and S-phase, respectively. 
Panels (a) and (e) are for the GL model, panels (b) and (f) are for the GPD model, panels (c) and (g) are for PhaseNet, and panels (d) and (h) are for EQT. These panels display probabilities larger than 0.1.}
\label{eventnumber}
\end{figure}

Second, we analysed the phase-detection performance in terms of phase probability. Because the number of phase detections changes depending on the probability threshold, we evaluated the performance without specifying the threshold. We focused on probabilities at least 0.1, discarding almost zero probabilities of the vast majority of the phase detections by these methods.  
The distributions of phase probabilities before and after the onset of the swarm imply that there are differences in performance depending on 
methods (Fig.~\ref{eventnumber}). To quantify these differences, we evaluated the area under the curve (AUC) of a receiver operating characteristic curve \cite{fawcett2006introduction, hand2012assessing} as follows: First, the label \mysingleq{before} was assigned to the probabilities that were yielded before the onset, whereas the label \mysingleq{after} was assigned to those after the onset. Next, a receiver operating characteristic curve was constructed for a binary classifier of \mysingleq{before} and \mysingleq{after} with various thresholds of probabilities, which were subsequently used for the AUC evaluation. 
For the P-phase, the AUC values for the GL model, the GPD model, PhaseNet, and EQT were 0.78, 0.66, 0.76, and 0.76, respectively. For the S-phase, the AUC values were 0.91, 0.79, 0.67, and 0.63, respectively, for the same models. These results imply that, for both phases, the GL model yielded the most separable probability distribution before and after the onset of the swarm. 

\subsection{Low-frequency earthquake in Japan}\label{LFEdetection}
Finally, we examined the detection performance for LFEs, which differ in type from conventional earthquakes, in which the low-frequency component (2-8 Hz) dominates in  the 
seismic wave \cite{ ide2007scaling}. From both the theoretical and observational perspectives, it is inferred that an LFE may be caused by Brownian motion throughout the area hosting the shear slip \cite{ide2008brownian, ide2018seismic}. In particular, LFEs have recently gained considerable attention to gain a better understanding of earthquake mechanisms and for their  possible connection to large earthquakes \cite{obara2002nonvolcanic, shelly2007non, peng2010integrated, kocharyan2021nucleation, kato2021generation}. Hence, it is particularly important to correctly detect LFEs, which provide valuable information for such research.
However, detecting LFEs is challenging because they manifest very weakly in waveforms with orders of magnitude below those of typical earthquakes \cite{thomas2021identification}. Moreover, owing to the low frequency of the waveform, a detection model trained using typical EQ data is not readily applicable for LFE detection \cite{thomas2021identification}. Hence, the current practice of LFE detection is based on human analysts making manual selections \cite{kurihara2021spatiotemporal} or on a matched-filter technique in which a new LFE is identified by cross-correlation analysis of waveforms with known LFEs \cite{shelly201715, kato2020detection}. In this study, as a novel attempt, we aimed to detect LFEs effectively using the proposed method’s flexible framework. 

In this experiment, we focused on the LFEs in the Tohoku region of Japan (latitude $37^{\circ}$N to $41^{\circ}$N and longitude $139^{\circ}$E to $142^{\circ}$E), as recorded in the JMA catalogue from 2015 to 2019 \cite{JMA2}. We evaluated whether our method could detect the catalogued LFEs. For better demonstration, we selected the LFEs from the JMA catalogue that met the following conditions:
\begin{itemize}
    \item Condition 1: There was no occurrence of other LFEs 10 min before and after a target LFE.
    \item Condition 2: In the same period as in Condition 1, there were no occurrences of conventional earthquakes with magnitudes of less than 3 within 100 km from the epicentre of the target LFE.
    \item Condition 3: Also, in the Condition 1 period, there were no occurrences of earthquakes with magnitudes larger than 3 in the Tohoku region.
\end{itemize}
These conditions allowed us to assume the absence of influences by other earthquakes  10 min before and after the target LFE. Furthermore, to be more rigorous, we retrieved the waveform 5 min before and after the target LFE.

We obtained waveform data from a high-sensitivity seismograph network (Hi-net) operated by the National Research Institute for Earth Science and Disaster Resilience (NIED), Japan \cite{okada2004recent}. Hi-net observation stations are densely situated throughout Japan with a 20-km mesh and routinely collect 3-component waveform data. 
First, the Hi-net station nearest  to the epicentre of the target LFE was identified. Subsequently, we extracted 10-min waveforms with three components 5 min before and after the occurrence of the target LFE. If a complete waveform was not available, the LFE was discarded. As a result, we obtained full waveforms for 446 LFEs.
No preprocessing was performed for these waveforms.  

\begin{figure}
    \centering
    \includegraphics[scale=1, trim=8mm 0mm 0mm 0mm]{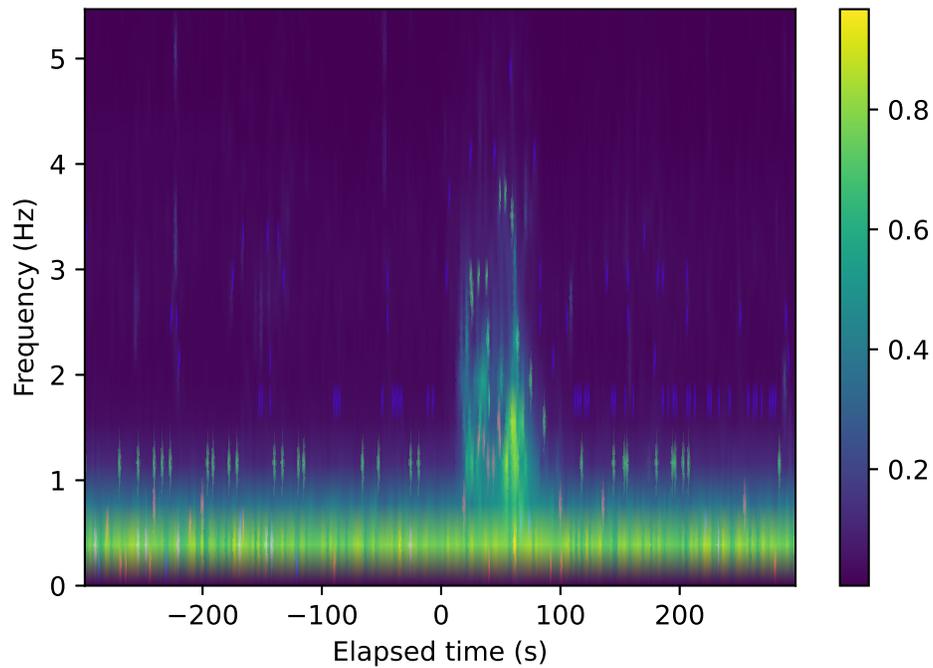}
    \caption{Mean dynamic spectrogram of the waveform’s up-down component for 446 LFEs in study. The horizontal axis denotes the time points in which the LFE’s origin time is set to 0 and the vertical axis, frequency (Hz). The colour denotes the mean intensity of the power. The spectrogram was generated as follows.
    First, for each LFE, the 10-min (600 s) up-down component of the waveform was standardised. Second, we generated a spectrogram for each LFE 
    using the Python function \mysingleq{scipy.signal.spectrogram} with the default setting (Tukey window with a shape parameter of 0.25; a segment length of 256 data points; no overlapping of segments) \cite{2020SciPy-NMeth}. Lastly, the mean spectrogram was evaluated for all 446 LFEs. 
    }
    \label{fig:spectrum}
\end{figure}

We applied the GL model, the GPD model, PhaseNet, and EQT to the waveform data to detect the P-phases of the LFEs.
For each method, we set the same detection threshold (0.5, 0.98, 0.5, and 0.3 for GL model, GPD model, PhaseNet and EQT, respectively) and window width (4 s, 4 s, 30 s and 60 s for GL model, GPD model, PhaseNet and EQT, respectively), as in the Bombay Beach swarm analysis described in Section~\ref{sectionbombay}. Because information on the exact timing
of the P-phase reaching an observation station was not available, we set the maximum time delay time to 100 s. In other words, we considered detection to be successful when the method in question detected the P-phase within 100 s after the LFE’s time of occurrence. This presumed delay time is based on the observation that high-intensity waveform power of between 1-4 Hz exclusively occurred in this period (Fig.~\ref{fig:spectrum}. For more details, please refer to the fourth paragraph of Section \ref{discussion}). Similarly, we considered detection to be false when the method detected the P-phase within 100 s before the LFE’s time of occurrence. 

\subsubsection*{Results}
\begin{figure}
\includegraphics[scale=0.35, trim=20mm 0mm 0mm 0mm]{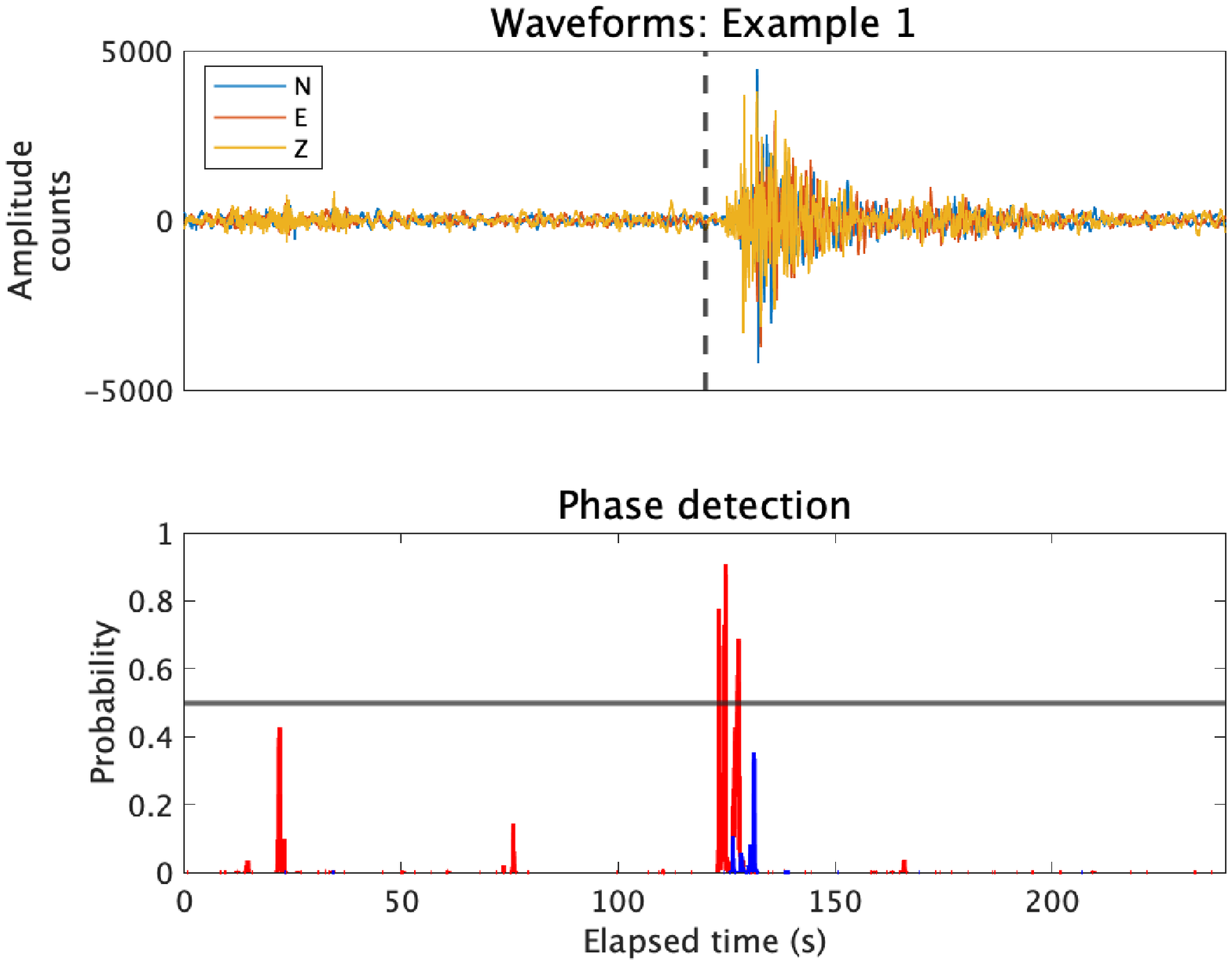}
\includegraphics[scale=0.35, trim=0mm 0mm 0mm 0mm]{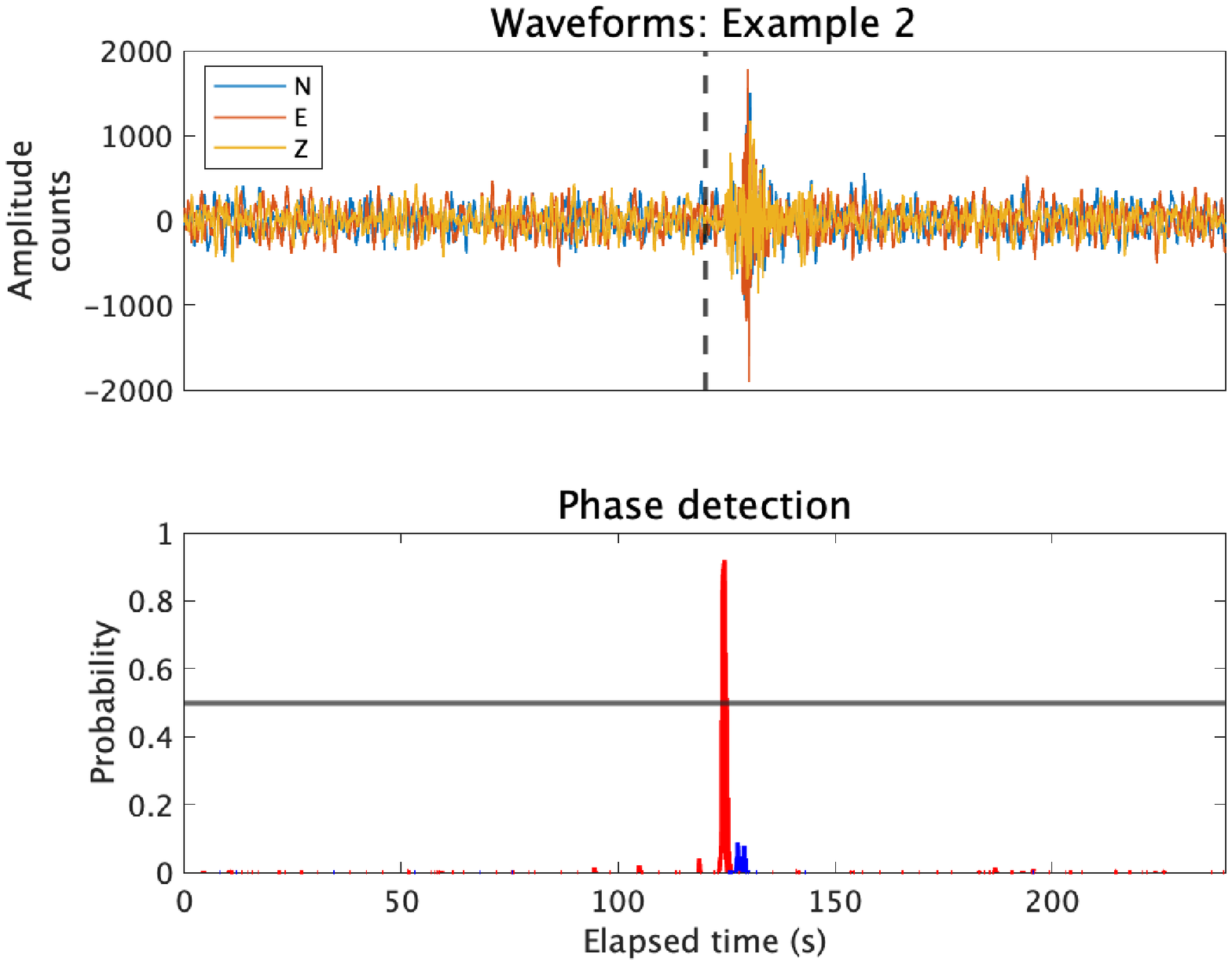}

\vspace{10mm}

\includegraphics[scale=0.35, trim=20mm 0mm 0mm 0mm]{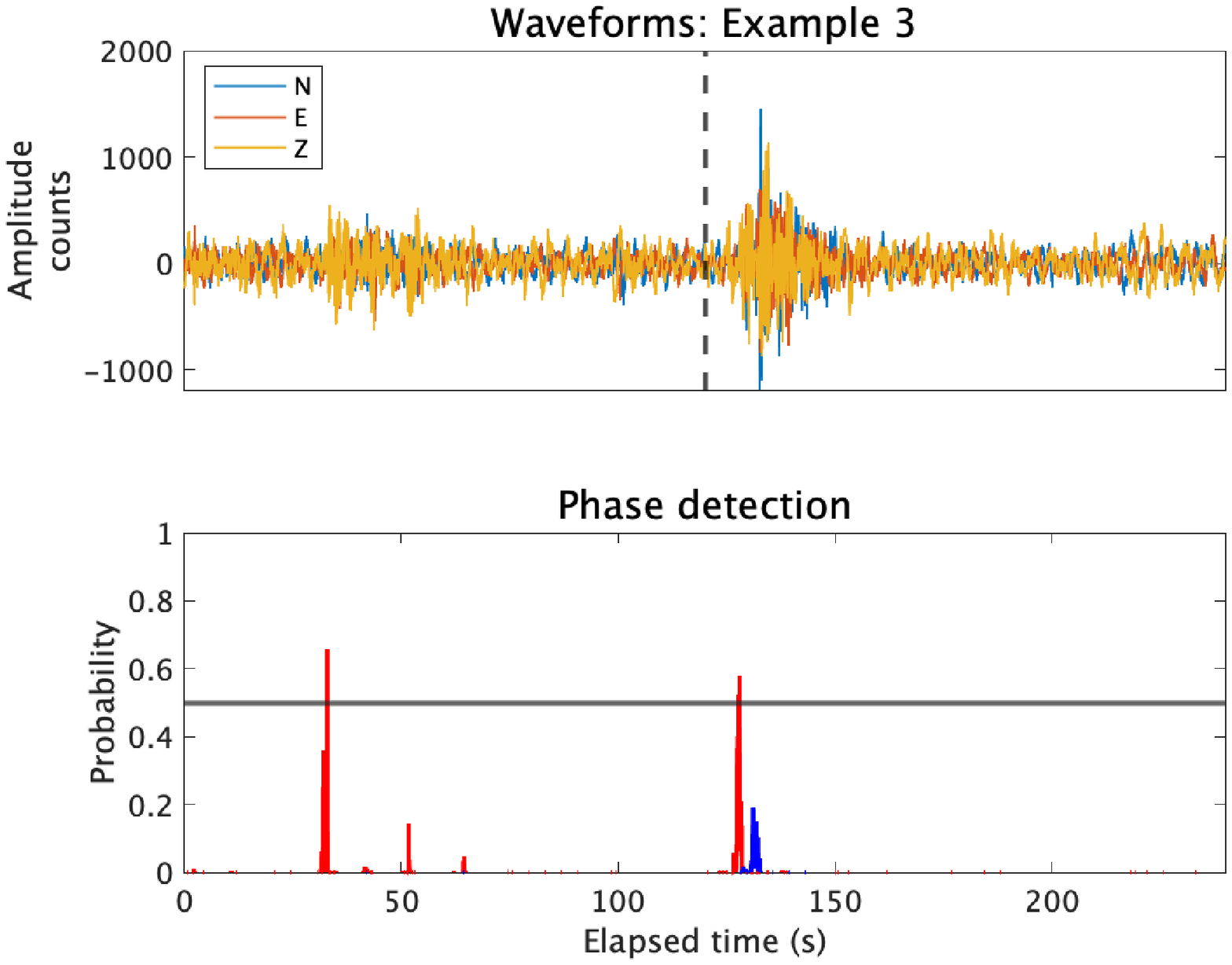}
\includegraphics[scale=0.35, trim=0mm 0mm 0mm 0mm]{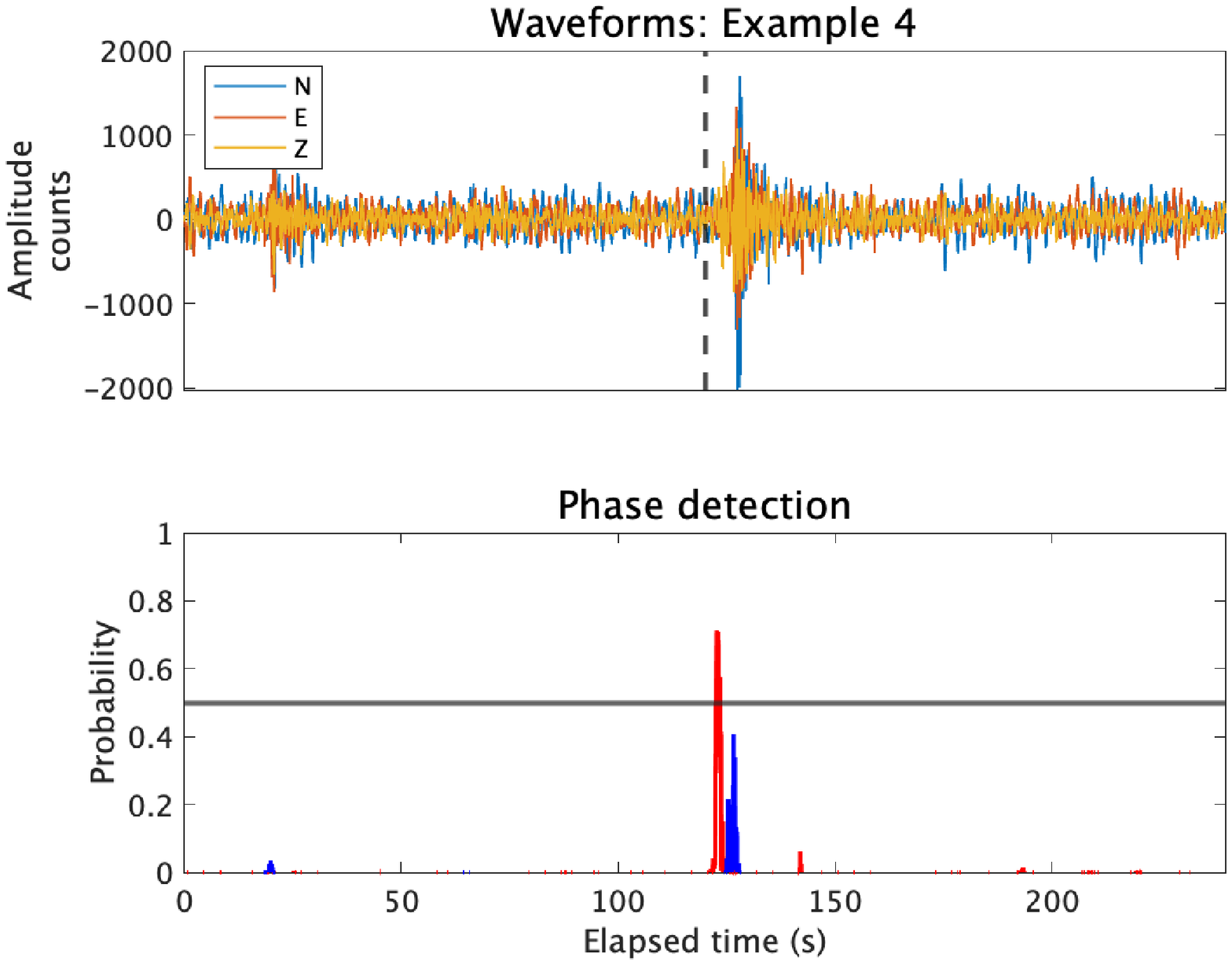}
\caption{Examples of LFE detection by the GL model. The 4-min (240 s) waveform of three components and phase detection probabilities are displayed for the upper sub-panel and lower sub-panel, respectively. For phase detection, the red line denotes a probability for P-phase, whereas the blue line is for S-phase. The horizontal black line denotes the threshold (0.5) for phase detection. The displayed examples were randomly sampled from those cases (78 cases) in which the GL model detected a P-phase within 30 s after the origin time . 
The origin time is centred, which is denoted by a dashed line in the upper panel. Note that there are some time lags between the origin time and P-phase detection, possibly due to the geometrical distance between the hypocentre and an observation station. 
}
\label{LFEexampleGL}
\end{figure}

The detection results obtained using these settings are summarised in Table~\ref{tabLFE}. The GL model correctly identified 94 out of 446 LFEs (21 $\%$), whereas the GPD model, PhaseNet, and EQT correctly identified  213 (48 $\%$), 38 (8.5 $\%$), and 94 (21 $\%$) LFEs, respectively (third column in Table~\ref{tabLFE}). For visual inspection, Fig.~\ref{LFEexampleGL} shows randomly sampled waveforms that were correctly detected by the GL model.
Conversely, the false detection cases were 27 (6.1 $\%$), 97 (21 $\%$), 42 (9.4 $\%$), and 57 (12 $\%$) for the GL model, GPD model, PhaseNet, and EQT (second column in Table~\ref{tabLFE}), respectively. In terms of the ratio of correct to false detections, the GL model performed the best (ratio=3.5), followed by the GPD model (2.2), EQT (1.6), and PhaseNet (0.90)  (fourth column in Table~\ref{tabLFE}).

\begin{table}
 \caption{Summary of P-phase detections for the LFE 100 s before and after the origin time. The column \mysingleq{Before-segment} denotes the results for the time segment 100 s before LFE occurrence, whereas \mysingleq{After-segment} denotes the 100 s after an LFE occurred. The numbers in parenthesis denote percentages of detection over the total number of LFEs (446). The column \mysingleq{After/Before} denotes the ratio of detections before and after the origin time (i.e., the values in column \mysingleq{After-segment} over the values in column \mysingleq{Before-segment}). The methods \mysingleq{adapted GL} denote adapted GL models in which sub-model L2 was trained using the contaminated data with synthetic LFE waveforms  in Eq.(\ref{LFEcontaminated}). 
 The frequency band for the band-pass filter to generate synthetic LFE is indicated in parentheses after the method name \mysingleq{adapted GL} in the first column of the table.}
 \vspace{-5mm}
  \footnotesize
    \begin{center}
    \begin{tabular}{|c|c|c|c|}
    \hline
   Method & \textbf{Before-segment}  & \textbf{After-segment} & \textbf{After/Before (Ratio)}\\
    \hline
    GL & 27 (6.1 $\%$) & 94 (21 $\%$) & \textbf{3.5}\\ 
    GPD & 97 (21 $\%$) &  213 (48 $\%$)  &  \textbf{2.2}\\
    PhaseNet & 42 (9.4 $\%$) & 38 (8.5 $\%$) &  \textbf{0.90}\\
    EQT & 57 (12 $\%$) & 94 (21 $\%$) &  \textbf{1.6} \\
     adapted GL (1-2Hz) &  1 (0.2 $\%$) & 22 (4.9 $\%$) &  \textbf{22} \\ 
     adapted GL (2-4Hz) &  3 (0.7 $\%$) & 44 (9.9 $\%$) &  \textbf{15} \\ 
     adapted GL (4-6Hz) &  6 (1.3 $\%$) & 32 (7.2 $\%$) &  \textbf{5.3} \\
     adapted GL (6-8Hz) &  8 (1.8 $\%$) & 17 (3.8 $\%$) &  \textbf{2.1} \\
    \hline
    \end{tabular}
    \end{center}
    \label{tabLFE}
\end{table}

\subsubsection*{Adapted models}
In the next experiment, we adapted the GL model specifically to detect the LFE P-phase. As in Eq.(\ref{probdef}), the GL model consisted of three sub-models: G, L1, and L2. Among these sub-models, we retrained the L2 model using synthetic data. 
To mimic the LFE waveform, we contaminated the training waveform data (i.e., SCSN data) with synthetic LFE waveforms, as follows:
\begin{eqnarray}
    \bb{x}^{(p)} \leftarrow 0.5 \times \bb{x}^{(p)} + 0.5 \times \bb{s},
    \label{LFEcontaminated}
\end{eqnarray}
where $\bb{x}^{(p)}$ denotes an up-down component of the P-phase waveform, and $\bb{s}$ is the synthetic waveform of the LFE. We generated the synthetic waveform $\bb{s}$ based on the Brownian motion model in \cite{ide2008brownian}, which was subsequently band-pass filtered (for more details, please refer to Appendix~\ref{appenLFE} and Fig.~S2 in the Supplementary Material). We considered the following bands for the band-pass filters: 1-2 Hz, 2-4 Hz, 4-6 Hz, and 6-8 Hz. These low-frequency bands presumably reflect the dominant frequency of the LFE. 
Using these contaminated data (with a sample size of 150000), we retrained the L2 model, which was incorporated into a new GL model in Eq.(\ref{probdef}) (hereafter referred to as the \mysingleq{adapted GL model}); the remainder of the sub-models, including the G and L1 models, remained unchanged.  

The performance results of the adapted GL models are summarised in Table~\ref{tabLFE} (the sixth to the ninth rows).  
The number of correct detection cases was 22 (4.9 $\%$), 44 (4.9 $\%$), 32 (7.2 $\%$), and 17 (3.8 $\%$) respectively for the 1-2 Hz, 2-4 Hz, 4-6 Hz, and 6-8 Hz band-pass filters. 
In contrast, the false detection rates were 1 (0.2 $\%$), 3 (0.7 $\%$), 6 (1.3 $\%$), and 8 (1.8 $\%$) for the 1-2 Hz, 2-4 Hz, 4-6 Hz, and 6-8 Hz band-pass filters, respectively. 
Concerning the ratio of correct and false detections, the adapted model with 1-2 Hz band-pass filters performed the best (ratio=22), followed by the adapted models with the 2-4 Hz (15), 4-6 Hz (5.3), and 6-8 Hz (2.1) band-pass filters. For comparison with the GL model, the adapted models with the 1-2 Hz, 2-4 Hz, and 4-6 Hz band-pass filters significantly outperformed the GL model.

\begin{figure}
    \centering
    \includegraphics[scale=0.18, trim=120mm 0mm 0mm 0mm]{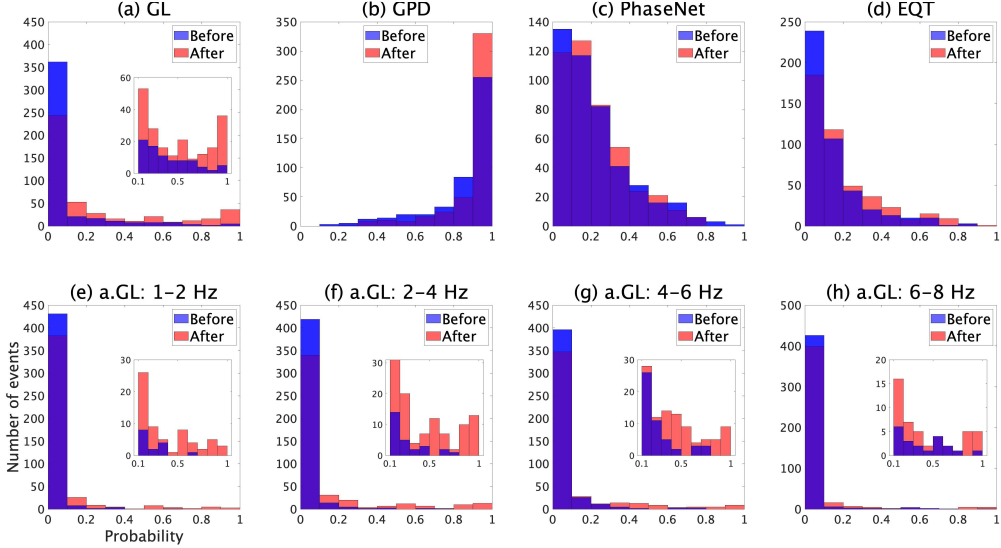}
    \caption{The maximum probabilities for P-phase detection before and after the LFE’s origin time. Each panel denotes a histogram for the number of maximum P-phase probabilities in the 100-s period before the LFE (blue bar) and in the 100-s period after the LFE (red bar). The horizontal axis denotes the maximum probability for P-phase detection, and the vertical axis denotes the number of LFE events (out of 446 events). 
    Panels (a)-(h) are for the GL model, the GPD model, PhaseNet, EQT, and adapted GL models, respectively, with the 1-2 Hz, 2-4 Hz, 4-6 Hz, and 6-8 Hz band-pass filters. For Panels (a), (e), (f), (g), and (h), 
    the inset figure shows the same histogram as the main figure,
    which focuses on a maximum probability larger than 0.1.
    }
    \label{fig:LFEseparationAUC}
\end{figure}

\begin{figure}
    \centering
    \includegraphics[scale = 0.6]{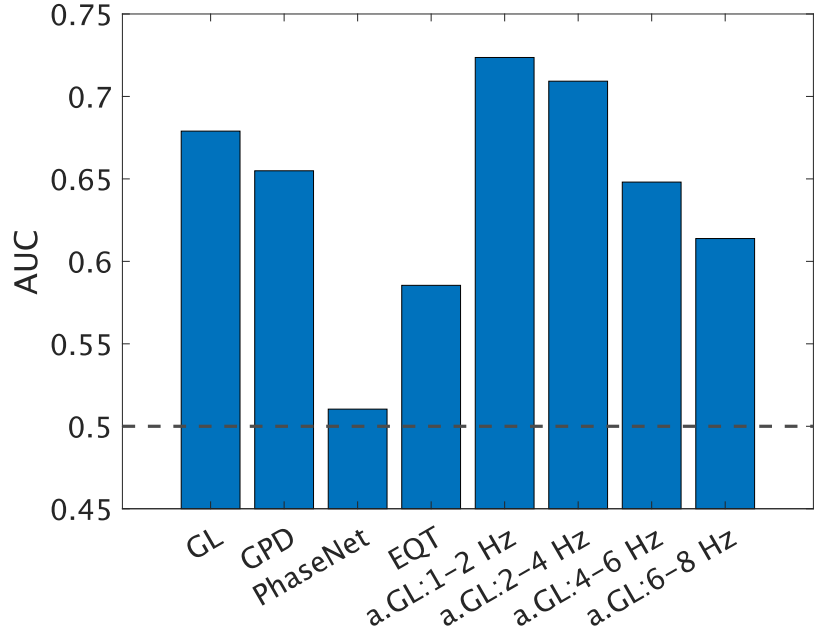}
    \caption{The AUC evaluation of the detection differences before and after the LFE .
    For each LFE (out of 446), we 
    evaluated the maximum probability of P-phase detection before and after the origin time. As a result, there are 446 maximum probabilities for the time segment \mysingleq{before} and the time segment \mysingleq{after}, respectively. The distributions of these maximum probabilities are displayed in  
    Fig.~\ref{fig:LFEseparationAUC}. 
    For these maximum probabilities of P-phase detection, the AUC is evaluated for binary time segment classification \mysingleq{before} and \mysingleq{after}, which, between two time segments, measures the separability of the maximum probabilities. 
    The horizontal axis denotes detection methods, whereas the vertical axis AUC value. 
    The dashed line denotes the chance level of the AUC value, i.e., 0.5. Abbreviations for adapted GL model: \mysingleq{a.GL:1-2 Hz}, adapted GL model with the 1-2 Hz band-pass filter; \mysingleq{a. GL: 2-4 Hz}, adapted GL model with the 2-4 Hz band-pass filter; \mysingleq{a. GL: 4-6 Hz}, adapted GL model with the 4-6 Hz band-pass filter; \mysingleq{a. GL: 6-8 Hz}, adapted GL model with the 6-8 Hz band-pass filter.}
    \label{fig:LFEAUC}
\end{figure}

Finally, we analysed the performance of the P-phase detection regarding detection probability. We evaluated the maximum probability of the P-phase before and after the LFE’s origin time. In  this manner, two probabilities were assigned for a single waveform. The distributions of these probabilities are shown in Fig.~\ref{fig:LFEseparationAUC}.
It is observed that the GL model and the adapted GL models with 1-2 Hz and 2-4 Hz band-pass filters separate the two segments to some extent. However, PhaseNet and EQT did not discriminate well between the two segments. Notably, for the GPD model, an effective separation between two segments is mainly observed at a probability larger than 0.9 but is not clearly visible in Fig.~\ref{fig:LFEseparationAUC}.  Furthermore, to summarise separability based on detection probabilities, we evaluated the AUC (Fig.~\ref{fig:LFEAUC}). The adapted GL model with a band-pass filter of 1-2 Hz performed the best (AUC=0.72), followed by the adapted GL model with a band-pass filter of 2-4 Hz (0.71), GL model (0.68),  GPD model (0.65), adapted GL model with a band-pass filter of 4-6 Hz (0.65), adapted GL model with a band-pass filter of 6-8 Hz (0.61), EQT (0.59), and PhaseNet (0.51).  

\section{Discussion} \label{discussion}
We propose a novel method for seismic-phase detection. The proposed method is an extension of the existing GPD method that considers local waveform information as well as global information. The method is based on deep learning using CNN architectures, which separately model the global and local representations of the waveform. The novelty of our method is that it explicitly focuses on both local and global representations, and subsequently combines them as a probability product. To our knowledge, there has been no neural network method that uses such a separate learning strategy for seismic-phase detection. For other application fields, the separate learning strategy has been used for pedestrian detection \cite{wang2018pcn} 
and action detection \cite{peng2016multi, tu2018semantic}. The framework of our method is similar to the attention mechanism adapted by the EQT method \cite{mousavi2020earthquake}, which implicitly determines the focal data points in a data-driven manner. The attention mechanism determines the attention weights for each instance (i.e., a waveform in our context); hence, these weights are not necessarily common to different instances. Moreover, the interpretability of the results yielded by the attention mechanism remains controversial owing to the uncertainty of the relationships between the attention weights and model outputs \cite{jain2019attention, niu2021review}. 
In contrast, we explored prior knowledge by applying a multiple clustering method to waveform data in which the analyst’s pick for the P- and S-phases is centred. It was implied that the first and  second halves of the data points convey different information on the seismic phases. This prior knowledge provides an explicit and consistent framework for modelling the local structure of a waveform, which in turn enables robust phase detection and flexible modelling of local information. 

The classification performance using the SCSN dataset demonstrates the robustness of our method. When the waveform data were partially contaminated with noise, the separate learning strategy of the proposed method performed excellently, outperforming the global representation method (e.g., for a noise proportion of 0.75 in the first half of data points, the accuracy of the GL model and G model was 0.88 and 0.65, respectively).
In particular, it was found that when there was a difference in noise type between the first and second halves of the data points, the proposed method significantly improved the performance of noise phase detection.  
This was because the performance of neither the L2 model nor the L1 model was influenced by the noise proportion in the first or second halves of the data points, which in turn contributed to the high performance of the GL model.  
These results suggest that the proposed method can potentially reduce the false-positive cases caused by noise. 

The robustness of the proposed method was verified by applying it to the 2016 Bombay Beach swarm. Our method identified a considerable number of seismic phases after the onset of the swarm, whereas it identified only a few cases before the onset. Overall, the proposed method outperformed the other methods, including the GPD model, PhaseNet, and EQT. This observation is further supported by the result of the AUC analysis of the probability distributions, which suggests that the proposed method outperforms the other methods irrespective of the detection threshold. Moreover, in this application, a marked difference between the GPD model and the proposed method is evident. In Fig.~\ref{bombayall2} (before the swarm onset), the GPD model yielded a large number of high probabilities for the P- and S-phases, whereas the proposed method yielded few. By setting a high probability threshold (0.98), the GPD model reduces false-positive cases. Nonetheless, the interpretation of such a large threshold is not straightforward. In contrast, the proposed method sets the threshold to 0.5, which facilitates the theoretical interpretation of phase probability. 

Moreover, the application of the proposed method  to LFE detection in Japan demonstrated the usefulness of the separation learning strategy. In this experiment, based on the Brownian motion model, we generated synthetic LFE waveforms, which were further bandpass filtered. Subsequently, we retrained the L2 model using contaminated data with synthetic LFEs.
This adapted procedure obtains better performance from the adapted models with 1-2 Hz and 2-4 Hz band-pass filters than the other methods, including the GL model, the GPD model, PhaseNet, and EQT. Overall, this result is consistent with the nature of the LFE waveform, in which the lower frequencies dominate. To clarify this, we performed a supplementary analysis to evaluate the mean dynamic spectrogram of 10-min waveforms of 446 LFEs (Fig.~\ref{fig:spectrum}). It was observed that before the origin time, a frequency band of less than 1 Hz (background noise) dominated. In contrast, a high-power intensity for 1-4 Hz was observed between the 0-100-s period, after the origin time. This suggests that an LFE occurred during this time with a characteristic frequency of 1-4 Hz. This frequency that dominated the waveform data may be attributed to the better performance of the adapted models with 1-2 Hz and 2-4 Hz band-pass filters. Notably, the P-phase detection of the test data (conventional EQ) by these adapted models deteriorates considerably (recall is 0.035 without noise contamination, Fig.~S3 in the Supplementary Material), which suggests that the observed good performance may be limited to the LFEs. Furthermore, theoretically, this application demonstrates that the proposed method has a specific form of transfer learning in which the learned model (conventional EQ) is adapted to a different context (LFE) by re-training a sub-model (i.e., the L2 model). This type of transfer learning is a unique feature of the proposed method owing to the separate learning strategy of the sub-models. Moreover, this application suggests the possibility of developing an LFE phase-detection model without using the LFE waveform, which is often not readily available with phase labels \cite{kurihara2021spatiotemporal}.  

Next, we discuss further theoretical aspects of the proposed method, which are useful for feature research on the development of phase-detection methods. The key idea of our method is based on the observations that the first and second halves of the waveform per se have sufficient information to discriminate among the P-phase, S-phase, and noise (Fig.~\ref{recall}). Existing detection methods can potentially improve their performance by incorporating local information as an additional component in the model. Second, the proposed method can be improved by designing completely different neural network architectures for local information. For local models, we simply adopted the same CNN architecture as the global model; however, it would be worth considering choosing a CNN or other architecture that is best suited for the local model. For instance, one may consider a Long Short Term Memory Network (LSTM) \cite{hochreiter1997long, sherstinsky2020fundamentals} for a L2 model, which can potentially capture different time-dependency among three phases. 
Third, one may generalise the binary power weight in Eq.(\ref{probdef}), which allows for real-valued power weight. The quality of information for classification may differ between the global and local models, which may in turn depend on the waveform data in question. Hence, it is a promising approach to determine these weights in a data-driven manner by setting a user-defined criterion. 

Finally, we discuss the limitations of the present study. We evaluated the performance of phase detection for the 2016 Bombay Beach swarm before the onset of the swarm by removing the detection results from 1 min after the earthquakes in the SCEDC catalogue. Nonetheless, we cannot rule out the possibility of non-catalogued earthquakes. 
The same limitation applies to the performance evaluation of low-frequency earthquakes in Japan. Moreover, in the latter case, there is another limitation to the performance analysis: We assumed that all LFEs were detectable by waveform data from a single station. This assumption may result in an underestimation of the detection capabilities of these methods. Further detailed analysis of the waveform could potentially overcome some of the limitations, which are worth further investigation.  

\section{Data availability}
Southern California Seismic Network data (SCSN) used for the training, validation and test is available at:
https://scedc.caltech.edu/data/deeplearning.html. The continuous data for Bombay Beach swarm is available (https://scedc.caltech.edu/ \\
index.html), whereas that for low-frequency earthquakes at: https://www.hinet.\\
bosai.go.jp/?LANG=en. Catalog of LFEs are available at: https://www.data.jma.\\
go.jp/eqev/data/bulletin/hypo.html (in Japanese). 

\section{Acknowledgements}
This research was supported by the MEXT Project for Seismology toward Research Innovation with Data of Earthquake (STAR-E), grant no. JPJ010217. The authors acknowledge the valuable discussions with scientists working on the  JST CREST research projects (grant nos. JPMJCR1761 and JPMJCR1763), Grant-in-Aid for Challenging Exploratory Research (grant no. 20K21785), Grant-in-Aid for Scientific Research (S) (grant no. 19H05662), ERI JURP, 2022-A-02, 2021-B-01, and 2022-B-06. The authors have no conflicts of interest to declare.

\bibliographystyle{plain}

\clearpage
\begin{appendices}
\section{Multiple clustering}\label{multi}
To reveal the multiple clustering structure of the data, we consider the multiple co-clustering method proposed in \cite{tokuda2017multiple, tokuda2018identification} (hereafter referred to as the \mysingleq{multiple clustering method}). The multiple clustering method is based on Gaussian mixture models, which assume that different object cluster solutions may be associated with a specific subset of features (Fig.~\ref{fig:multipleclustering}). Here, we refer to this subset of features as \mysingleq{part}. The method aims to identify the underlying parts and their object solutions that are probabilistically optimal under the non-overlapping constraint of feature partitioning (i.e., features are exclusively partitioned into these parts). In addition, the features in each part were clustered, yielding a co-clustering structure for the data. 
Assuming probabilistic independence
of cluster blocks in the inter- and intra-parts, 
such multiple data structures of clusters are modelled as a product of univariate Gaussian distributions. Furthermore, the method is based on nonparametric Bayesian statistics in which the number of parts and the number of clusters for both features and objects are automatically inferred. 
The method consists of three steps.
First, it partitions features into several parts, which works as a feature selection method for different
object cluster solutions. Second, it further partitions the features within a part that bundles similar features. Third, it partitions objects in each part, yielding several object cluster solutions. These three partitioning phases are performed simultaneously, yielding 
optimal feature partitioning and cluster solutions by fitting a univariate Gaussian distribution to each cluster block.
In this study, we applied this method to a waveform dataset in which an object corresponds to an instance of a waveform consisting of 400 data points (i.e., 400 features).

\renewcommand\thefigure{\thesection\arabic{figure}} 
\setcounter{figure}{0}
\begin{figure}
    \centering
    \includegraphics[scale=0.5, trim=20mm 20mm 0mm 0mm]{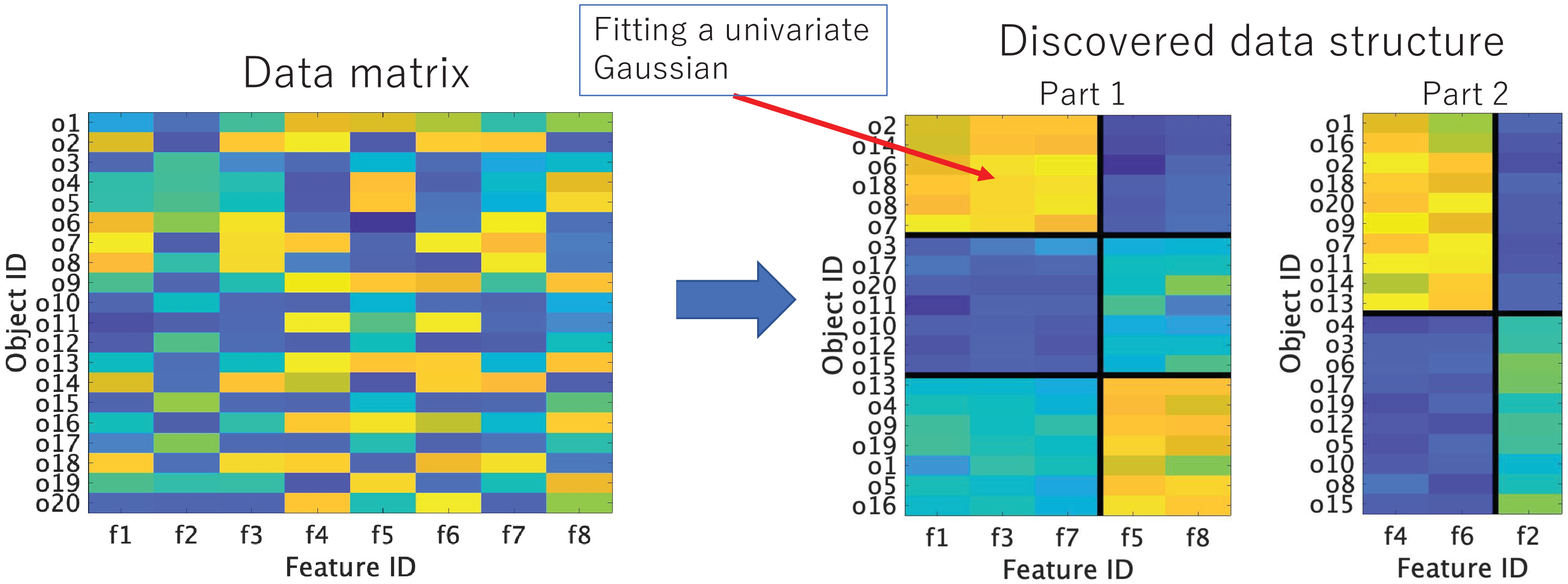}
    \caption{ Illustration of discovering the underlying data structure by the multiple clustering method. For these heatmaps, the horizontal axis denotes the feature ID (suffix \mysingleq{f}), whereas the vertical axis denotes the object ID (suffix \mysingleq{o}). Furthermore, the colour denotes the feature values. The multiple clustering method can recover the underlying data structure, which is illustrated in the two heatmaps on the right-hand side. 
    In this example, features are first partitioned into two parts, i.e., (f1, f3, f7, f5, and f8) and (f4, f6, and f2). Subsequently, the features in each part are further partitioned: (f1, f3, and f7) and (f5 and f8) for Part 1 and (f4 and f6) and f2 for Part 2. Lastly, objects are clustered for each part accordingly. In this illustration, the features and objects are sorted for those belonging to the same clusters  to be adjacent. The bold black lines denote boundaries between clusters for features and objects.
    The multiple clustering method simultaneously yields these feature partitions and object cluster solutions by  optimally fitting a univariate Gaussian distribution to each cluster block.}
    \label{fig:multipleclustering}
\end{figure}

\clearpage

\renewcommand{\theequation}{B.\arabic{equation}}
\setcounter{equation}{0}
\section{Recall/precision/accuracy}\label{recalldef}
Conventionally, recall and precision terminologies are used to evaluate the binary classification performance. In the present study, we extended this definition to multiclass classification performance. $n_{i, j} $ denotes the number of instances of the true phase $i$ that are detected as phase $j$ by a phase detection method. Here, we assume three phases in the context: $i, j \in \{\mysingleq{p}, \mysingleq{s}, \mysingleq{n}\}$.
We define the recall and precision of phase $i$ as follows:
\begin{eqnarray}
   \mbox{Recall} &=& \frac{n_{i, i}}{\sum_{j=1}^{3} n_{i, j}}\\
    \mbox{Precision} &=& \frac{n_{i, i}}{\sum_{j=1}^{3} n_{j, i}}.
\end{eqnarray}

\noindent Similarly, we define accuracy as follows:
\begin{eqnarray}
   \mbox{Accuracy} &=& \frac{\sum_{i=1}^3 n_{i, i}}{\sum_{i=1}^3 \sum_{j=1}^{3} n_{i, j}}.
\end{eqnarray}

\renewcommand{\theequation}{C.\arabic{equation}}
\setcounter{equation}{0}
\section{Synthetic LFE waveform} \label{appenLFE}
To generate a synthetic LFE waveform, we followed the Brownian walk model proposed by \cite{ide2008brownian}. Here, we summarise the model, including the specifications of the relevant parameters. 

First, we consider a circular fault with radius $r_t$ that changes dynamically with time $t$. We assume that the radius $r_t$ follows a differential equation:
\begin{eqnarray}
   dr_t = - \alpha dt + \sigma dB_t, 
    \label{drt}
\end{eqnarray}
where $dB_t$ is a random variable with a Gaussian distribution $N(0, dt)$, $\alpha$ is the damping coefficient, and $\sigma$ is the diffusion coefficient. Furthermore, assuming that shear slip occurs for this circular fault with a constant velocity $\nu_0$, the seismic moment rate $\dot{M}_0$ is given by
\begin{eqnarray}
    \dot{M}_0 = \mu \times \pi r_t^2 \times \nu_0, 
\end{eqnarray}
where $\mu$ is the rigidity. 
\noindent Hence, for a small value of $dt$, 
\begin{eqnarray}
  \no (\ddot{M}_0)^2 &\sim& 4\pi^2 \mu^2 \nu_0^2 \sigma^2 r_t^2/dt \\
   &\propto& r_t^2.
   \label{dotdot}
\end{eqnarray}
\noindent In \cite{ide2008brownian}, both intermediate-field and far-field were considered for S-wave generation. Here, for simplicity, we considered only the far field for P-wave generation. The first derivative of the P-wave displacement $u$ is given by equation (4.32) in \cite{aki2002quantitative} 
\begin{eqnarray}
    \dot{u}(t) = \frac{1}{4\pi \rho v_p^2 r}\bb{A}^{FP} \ddot{M}_0(t-r/v_p)
    \label{udot},
\end{eqnarray}
where $r$ denotes the distance from the source to the station. 
$\rho$ density, $v_p$ velocity of the P-wave, $\bb{A}^{FP}$ radiation patterns for the far-field P-waves. Our aim is to generate a waveform in a single component with an arbitrary amplitude (the amplitude is normalised later). Therefore, 
we further simplify Eq.(\ref{udot}) while ignoring the time delay.
\begin{eqnarray}
    \dot{u}(t) \propto  \ddot{M}_0(t) 
     \label{udot2}.
\end{eqnarray}
We consider that the waveform generated using Eq.(\ref{udot2}) mimics the LFE. $\ddot{M}_0(t)$ is generated based on Eq.(\ref{dotdot}), where the sign of $\ddot{M}_0(t)$ is determined by 
the sign of $\dot{M}_0(t+dt)-\dot{M}_0(t)$.
We set the relevant parameters as in \cite{ide2008brownian}. 
$\alpha=0.02~\mbox{s}^{-1}$; $\sigma=400~\mbox{m/s}^{1/2}$;
$dt=0.01~\mbox{s}$.
Using these formulations, we stochastically updated $r_t$ based on Eq.(\ref{drt}) for 2 s starting with $r_t=0$. For each time point, we evaluate $\dot{u}(t)$ in Eq.(\ref{udot2}), which constitutes a single LFE waveform. Because each time step $dt=0.01$, this procedure generates 200 data points of the LFE with a frequency of 100 Hz. Subsequently, we applied band-pass filters to the obtained waveform (1-2 Hz, 2-4 Hz, 4-6 Hz, and 6-8 Hz). We repeat this procedure to generate a synthetic waveform $\bb{s}$ for each instance of $\bb{x}^{(p)}$ in Eq.(\ref{LFEcontaminated}). Lastly, 
we normalised $\bb{s}$ such that the maximum absolute amplitude became identical to that of $\bb{x}^{(p )}$. 
\end{appendices}

\end{document}



\begin{flushleft}
\Large
\textbf{Supplementary Information for \mydoubleq{Seismic phase detection using multiple deep learning models for global and local representations of waveforms}}
\end{flushleft}
~~~\\
\noindent
\textbf{Tomoki Tokuda$^1$ \& Hiromichi Nagao$^{1, 2, *}$}\\
~~~~\\
\noindent
1. Earthquake Research Institute, The University of Tokyo, Japan\\
\noindent
2. Graduate School of Information Science and Technology, The University of Tokyo, Japan\\
\noindent
* Corresponding author: Hiromichi Nagao, nagaoh@eri.u-tokyo.ac.jp

\clearpage
\begin{figure}
\includegraphics[scale=0.7]{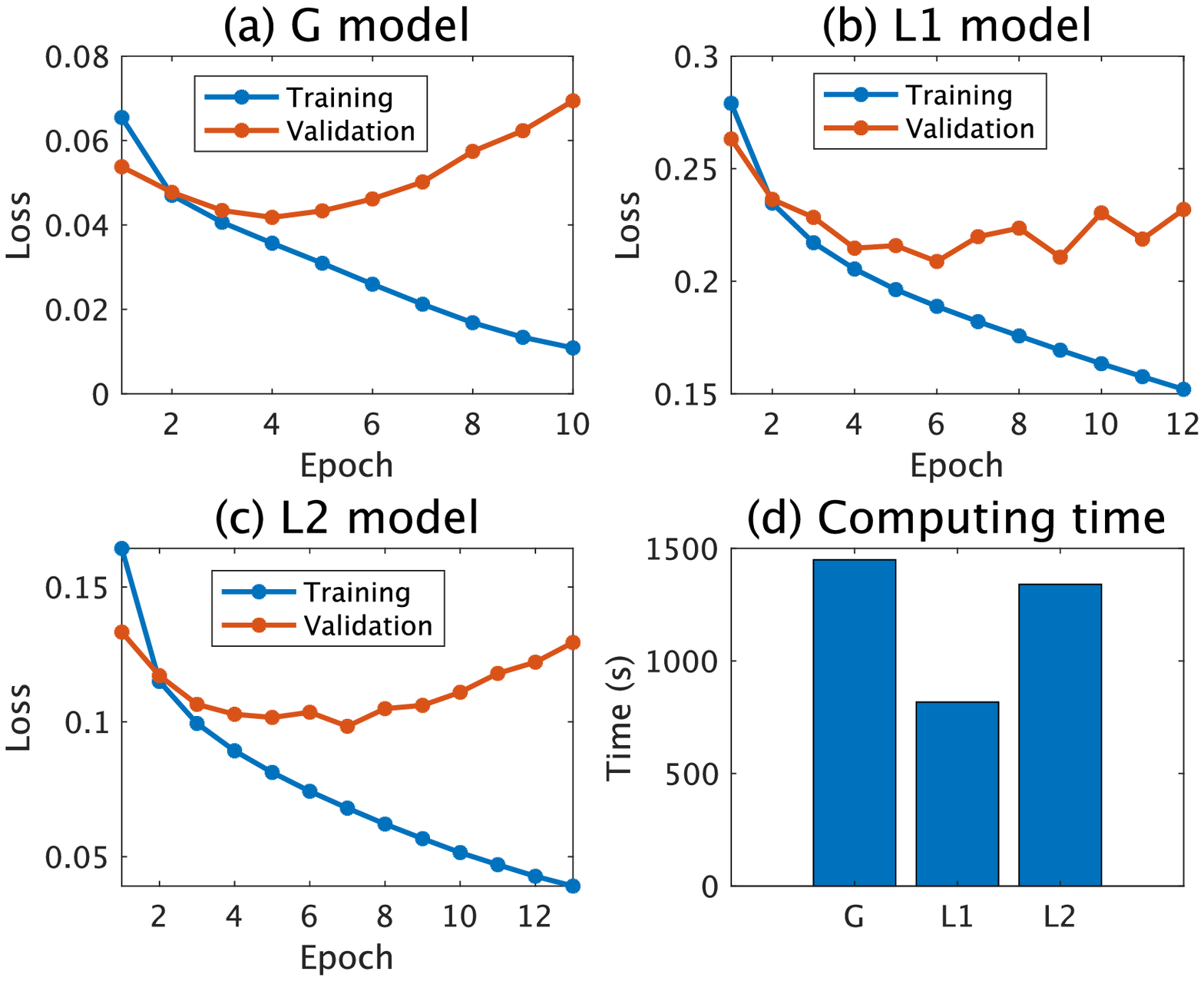}
\caption{Panels (a)-(c): Learning curves using all data points (\mysingleq{G model}), the first half of the data points (\mysingleq{L1 model}) and the second half of the data points (\mysingleq{L2 model}), respectively. Panel (d): Computation time (s) for training each model. We used four NVIDIA A100-PCIE-40GB graphical processing units.}
\label{Supp:epoch}
\end{figure}

\begin{figure}
\includegraphics[scale=0.35, trim=12mm 0mm -10mm 0mm]{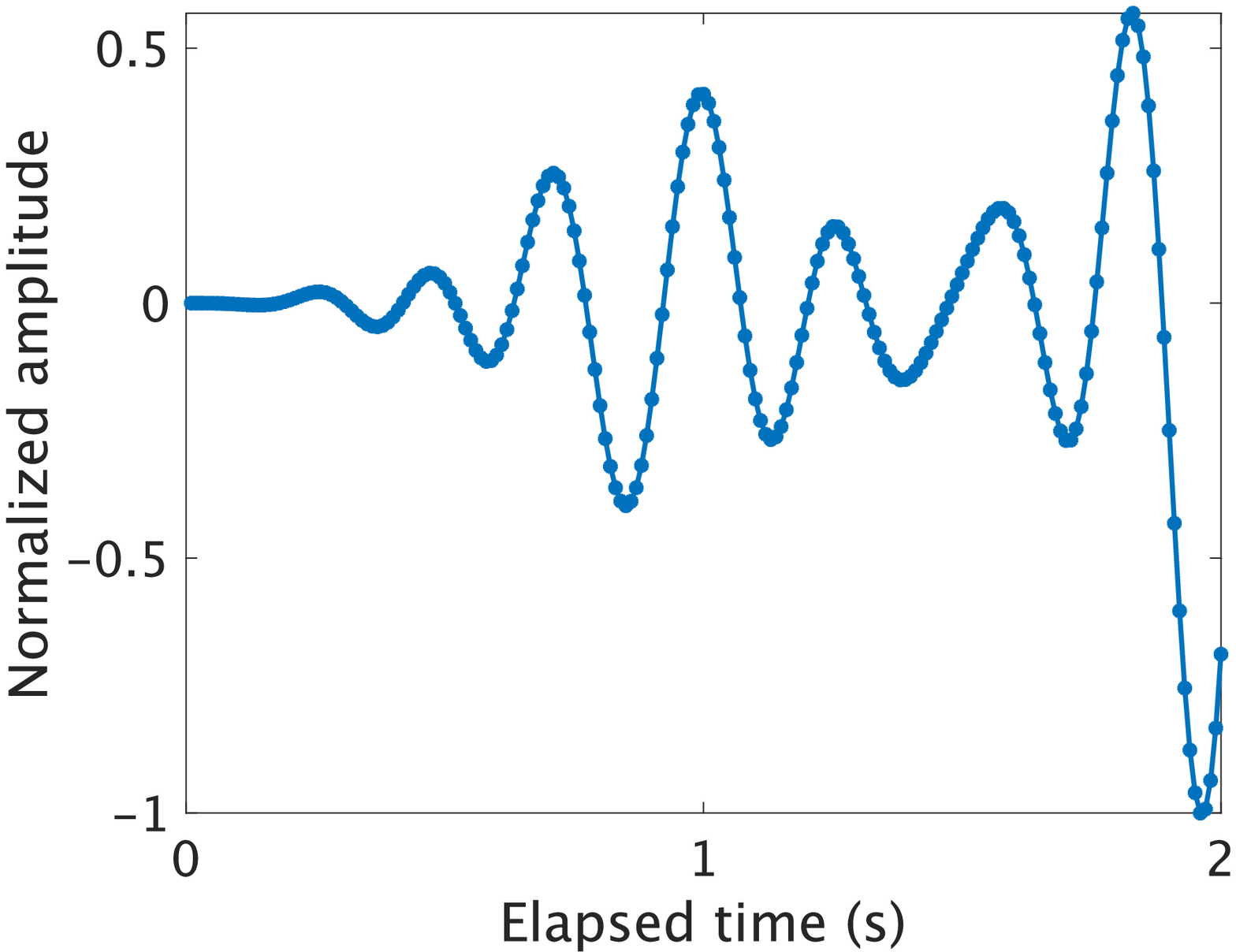}
\includegraphics[scale=0.35]{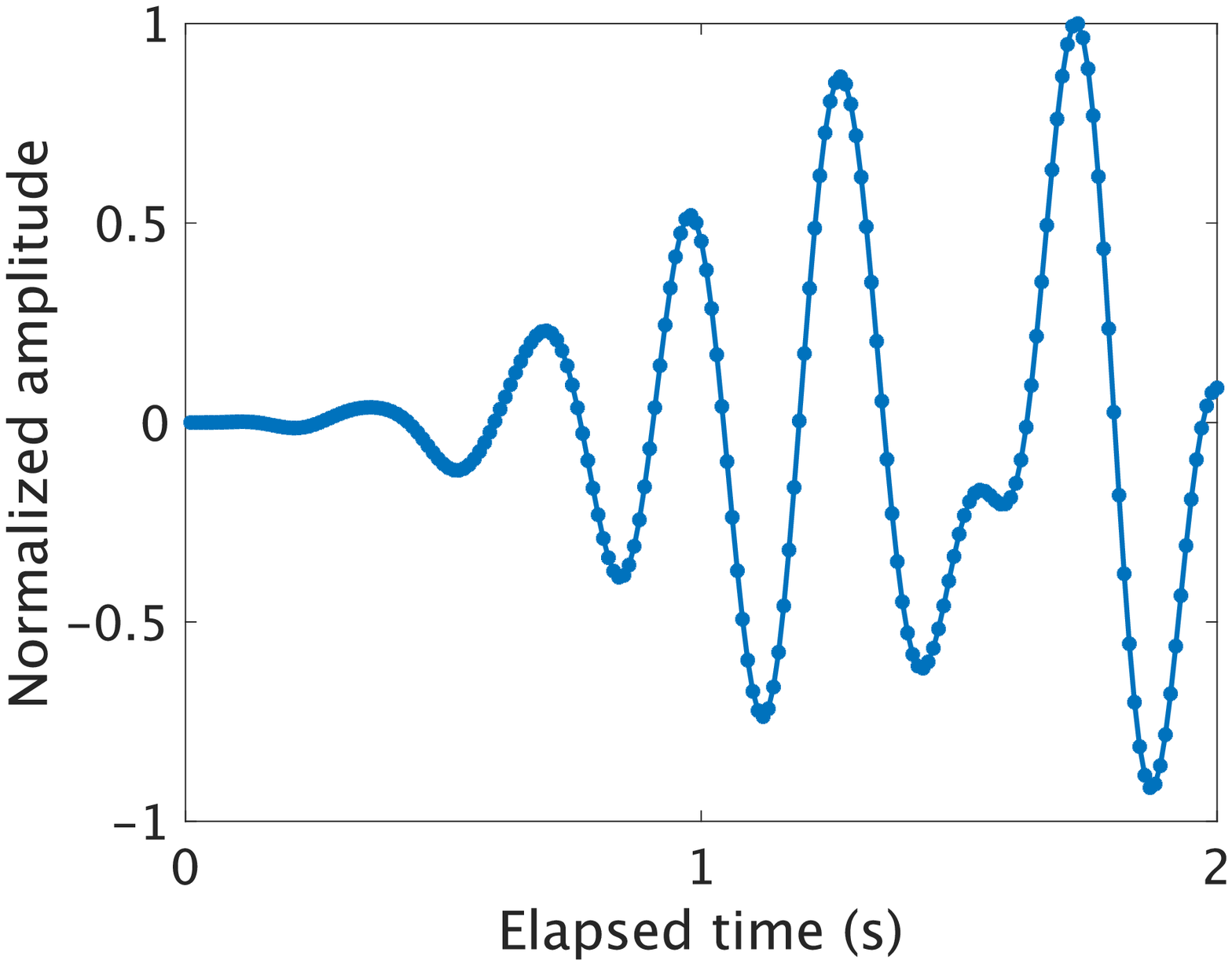}\\
\includegraphics[scale=0.35, trim=12mm 0mm -10mm -15mm]{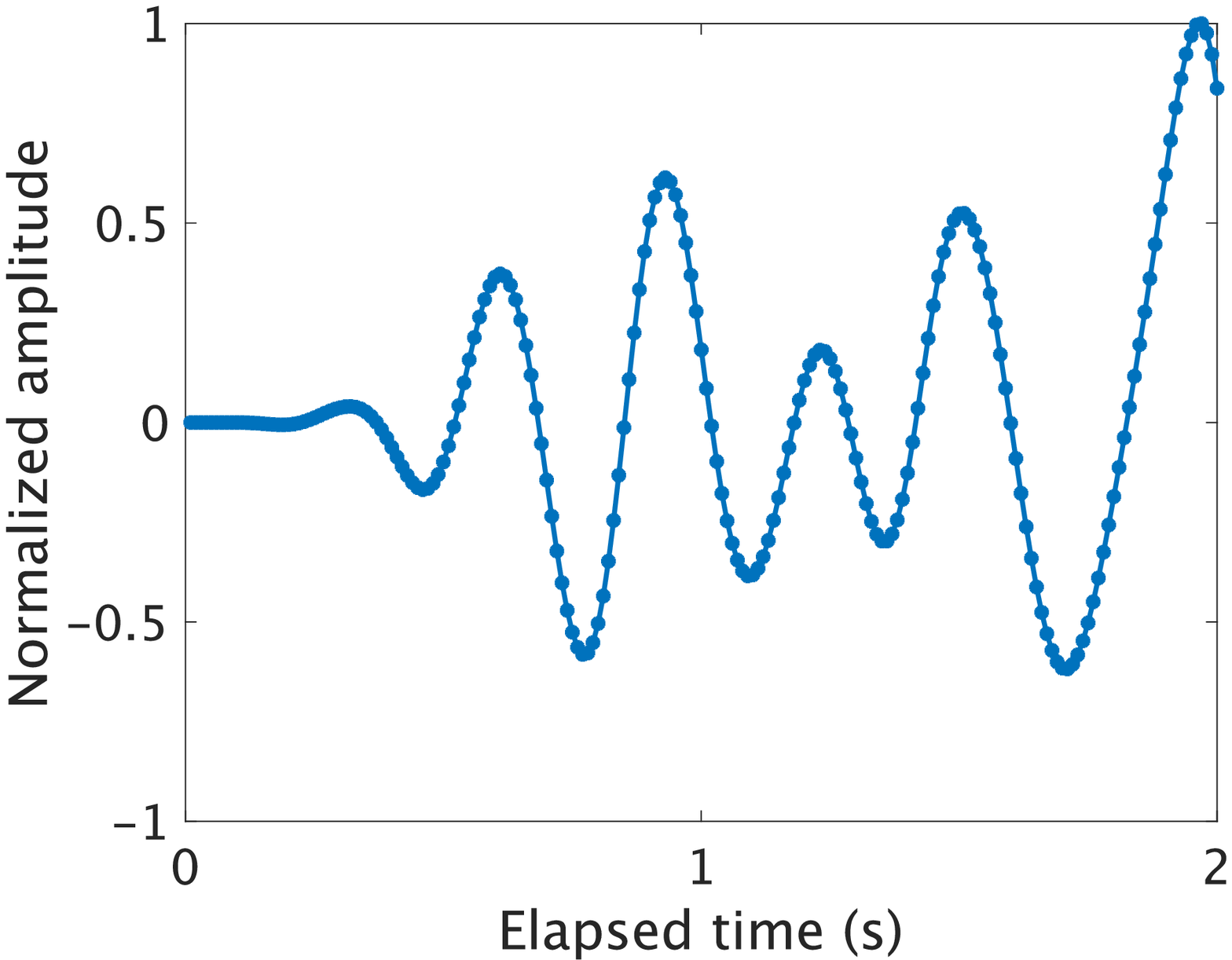}
\includegraphics[scale=0.35]{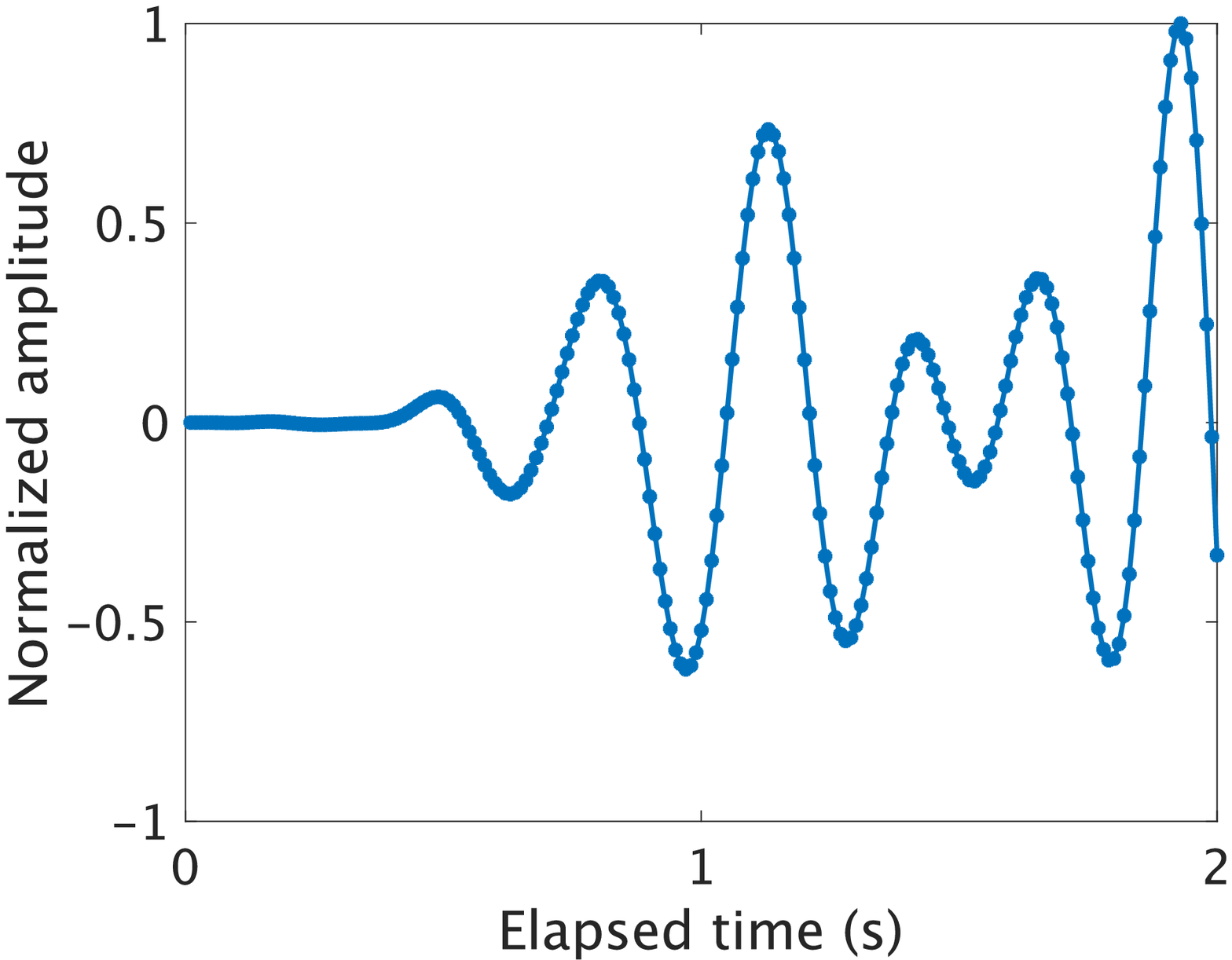}
\caption{Examples of synthetic LEF waveforms with the 2-4 Hz band-pass filter. Four randomly selected waveforms generated as described in Appendix C, are displayed. The horizontal axis denotes elapsed time and the vertical axis denotes amplitude normalised by the maximum absolute value.  
}
\label{Supp:LFE}
\end{figure}

\begin{figure}
\includegraphics[scale=0.4, trim = 25mm 0mm 0mm 0mm]{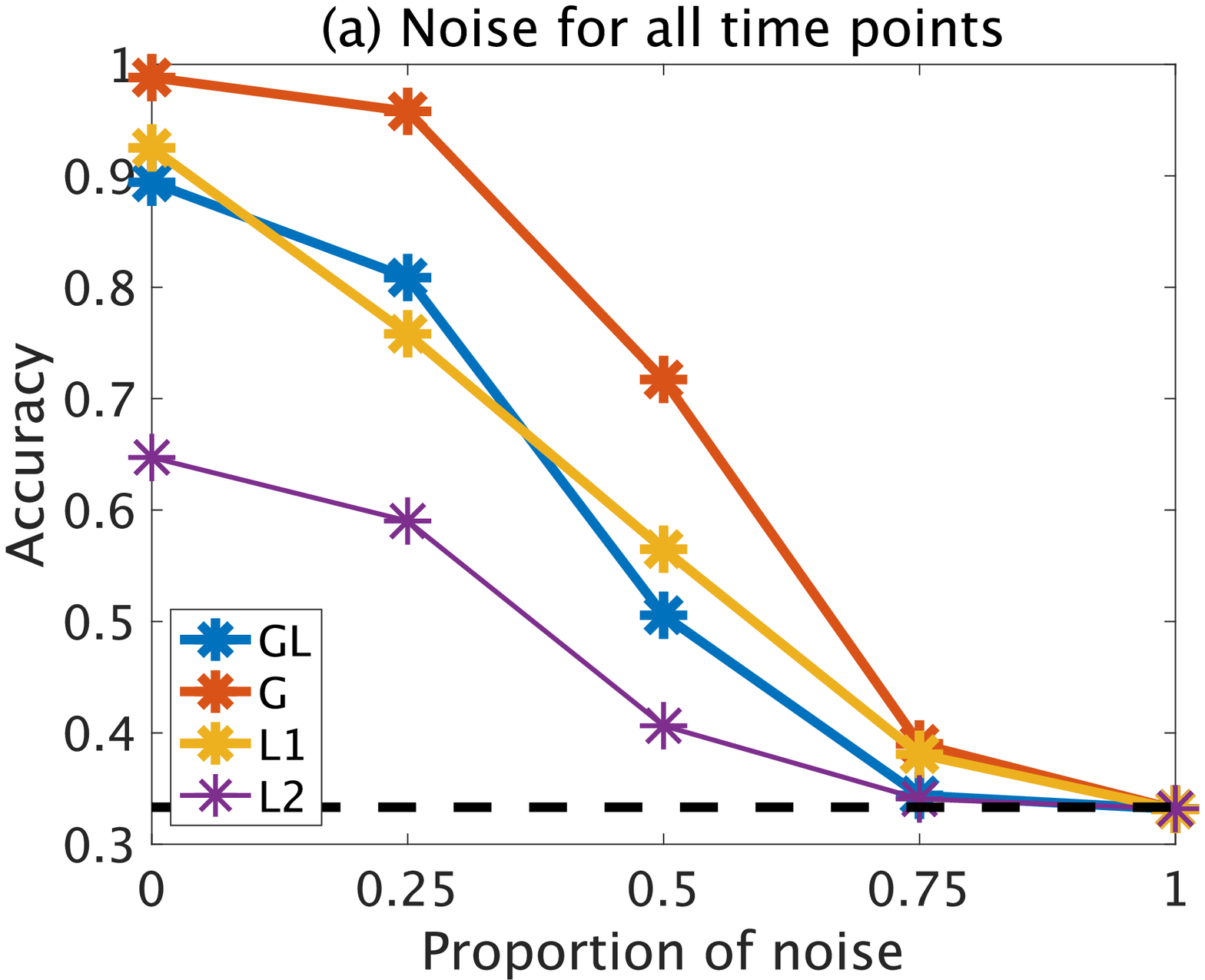}
\includegraphics[scale=0.4]{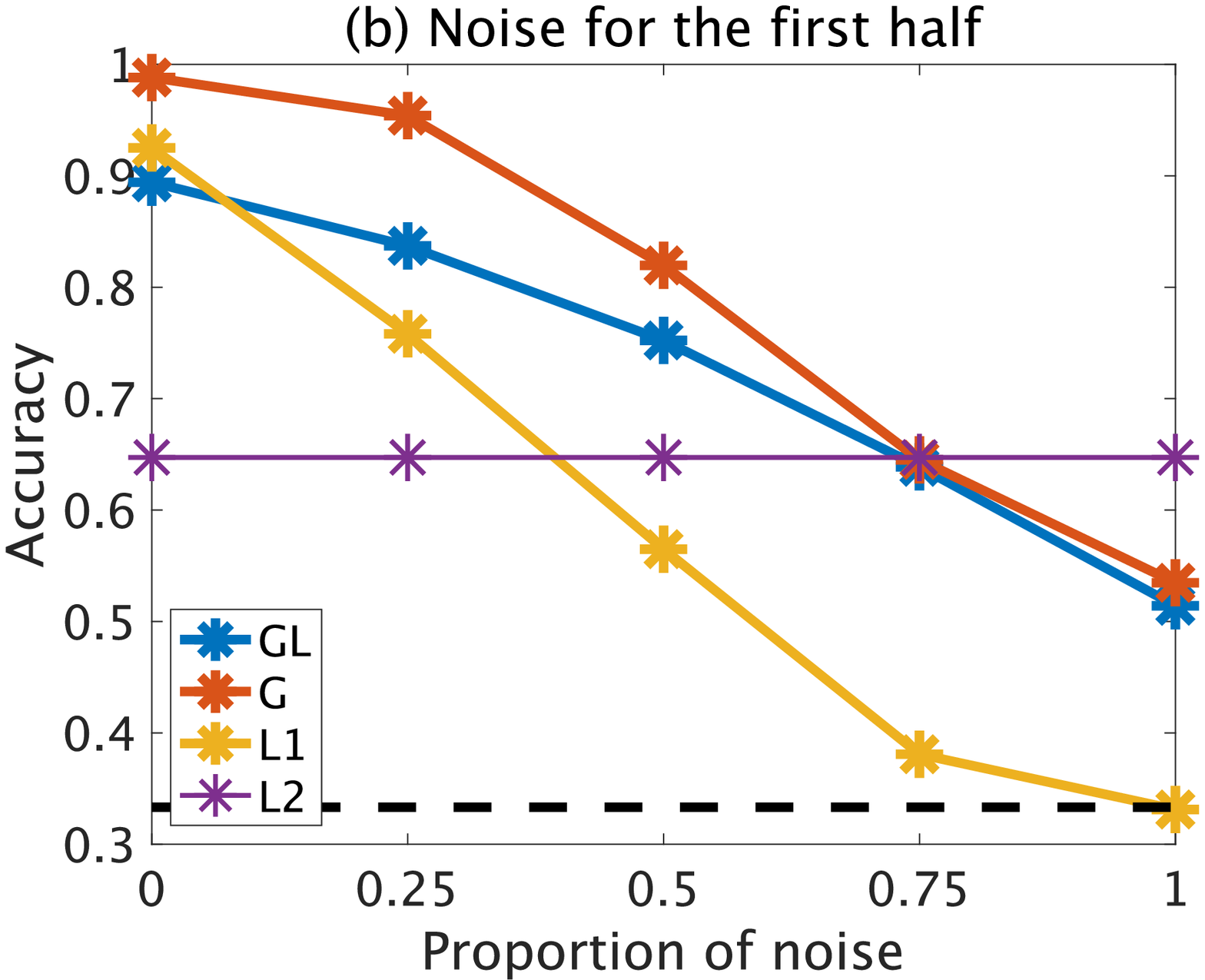}
\includegraphics[scale=0.4, trim = 25mm 0mm 0mm -10mm]{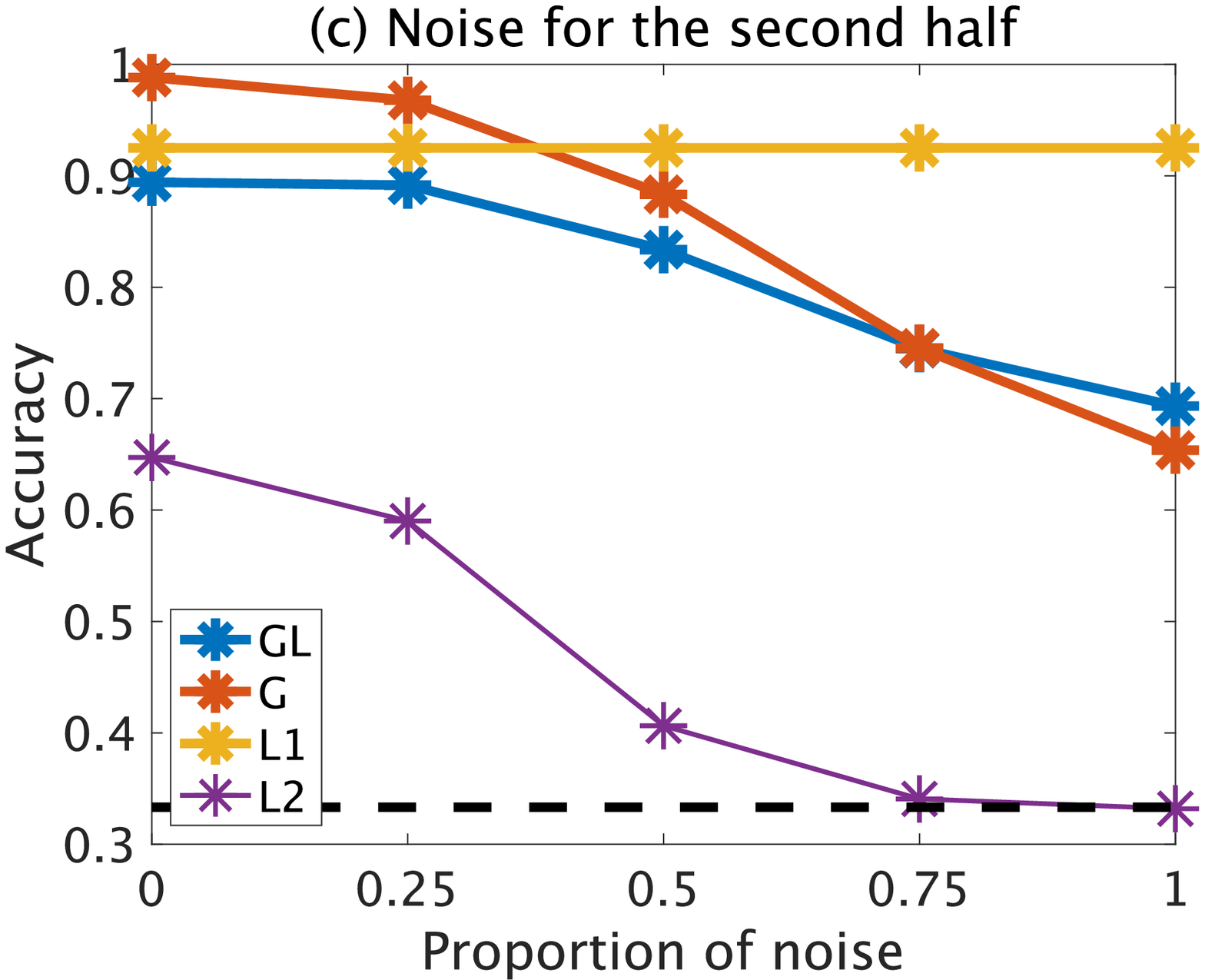}
\caption{Classification performance for 
the noise-contaminated test data when the L2 model was trained using the synthetic LFE data. The L2 model was trained as in Section 4.4 in the main manuscript in which the 2-4 Hz band-pass filter was applied to the synthetic LFE waveform. The remainder of the setting is the same as in Fig.~4 in the main manuscript. It is noted that the poor performance of the L2 model for the no-noise case (proportion of noise=0) is mainly due to the misclassification of the P-phase with the recall being 0.035. 
}
\label{Supp:performance}
\end{figure}

\clearpage

\begin{table}
\caption{Agreement of the cluster solutions among global representation and local representations in terms of the adjusted Rand index (ARI). ARI was evaluated for each pair of the cluster solution, which is shown by an upper triangular matrix in the table (diagonal elements are omitted). In row and column, \mysingleq{Global} denotes a cluster solution of the global representations (using all data points), whereas \mysingleq{Part 1-4} denotes cluster solutions of local representations yielded by the multiple clustering method (Fig.~2d in the main manuscript). 
ARI gives one when two clustering solutions completely match, whereas zero is given when two match only at a random level.}
\vspace{1.5mm}
\begin{tabular}{|c|ccccc|}
\hline
        &  Global & Part 1 & Part 2 & Part 3 & Part 4 \\
    \hline
 Global &     & 0.18  &  0.067  &  0.033  &  0.056 \\
Part 1  &   &     &   0.038 &   0.024 &   0.039 \\
Part 2  &  &    &       &  0.045  &  0.030 \\
Part 3  &    &   &     &       & 0.037 \\
Part 4  &     &  &     &     &    \\
\hline
\end{tabular}
\label{Supp:clustering}
\end{table}

\begin{table}
\caption{Classification performance for the test data by GL, G, L1, and L2 models. In the table, three phases, the P-phase, S-phase, and noise are denoted as \mysingleq{P}, \mysingleq{S}, and \mysingleq{N}, respectively. 
The phases in the vertical direction denote true phases (\mysingleq{True}), whereas those in the horizontal direction denote phases detected by a detection model (\mysingleq{Model}). The numbers in the table denote the number of instances of waveforms. The bold numbers denote the number of instances that were agreed between the true phase and the detected phase by a detection model.}
\vspace{-5mm}
\begin{flushleft}
    \begin{tabular}{|c|ccc||ccc|}
    \hline
    & \multicolumn{3}{c||}{GL} & \multicolumn{3}{|c|}{G} \\
    \hline
True\textbackslash Model  & P   & S & N &  P   & S & N \\
P & \textbf{156998} & 606  & 2211 & \textbf{157450} & 737 & 1628 \\
S & 524 &\textbf{157199}& 1151& 590 & \textbf{157307} & 977 \\
N & 817 & 1047&\textbf{156822} & 719 & 1168 & \textbf{156799}\\
     \hhline{|=======|}
    & \multicolumn{3}{c||}{L1} & \multicolumn{3}{|c|}{L2} \\
    \hline
True\textbackslash Model  & P   & S & N &  P   & S & N \\
P & \textbf{143534} & 7954  & 8327 & \textbf{154632} &  1317  & 3866 \\
S & 8467 &\textbf{148054}& 2353&  1055 & \textbf{154865} & 2954 \\
N &  6047 & 2579&\textbf{150060} & 2436 & 3394 & \textbf{152856}\\
    \hline
    \end{tabular}
    \end{flushleft}
    \label{suppli-allclassresults}
\end{table}

\begin{table}
\caption{ Classification accuracy performance for the noise-contaminated test data. The numbers in the table denote the mean values and standard deviations (in parentheses) for the GL, G, L1, and L2 models. The results are summarised with respect to the noise level and locus of noise contamination: \mysingleq{All} denotes contamination of all data points; \mysingleq{First half} contamination of the first half of the data points; \mysingleq{Second half} contamination of the second half of the data points.}
\vspace{-5mm}
\begin{flushleft}
\begin{tabular}{|ll|ccccc|}
\hline 
 \multicolumn{2}{|l|}{\textbf{Noise level}}   & \textbf{0} & \textbf{0.25} & \textbf{0.5} &  \textbf{0.75} &\textbf{1} \\
 \hline 
All & GL &  0.99 (0.0034) &    0.96 (0.0060) &   0.65 (0.014) & 0.37 (0.015) &   0.33 (0.015) \\
  & G  &   0.99 (0.0035)  & 0.96 (0.0057) & 0.72 (0.016) & 0.39 (0.015) &  0.33 (0.015)\\
  & L1 &  0.93 (0.0086) & 0.76 (0.013) &  0.56 (0.016) & 0.38 (0.016) & 0.33 (0.015)  \\
  & L2  &   0.97 (0.0050) &  0.88 (0.0096) &   0.52 (0.017) &  0.35 (0.016) &   0.33 (0.015) \\
  \hline
  First & GL & 0.99 (0.0034)&   0.98 (0.0041) &    0.95 (0.0064) &    0.88 (0.0091) &    0.75 (0.013)
 \\
  half & G &   0.99 (0.0035) & 0.95 (0.0063)  &  0.82 (0.013) &    0.65 (0.016) &    0.53 (0.015)\\
  & L1 &    0.93 (0.0086) &    0.76 (0.013) &    0.56 (0.016) &   0.38 (0.016) &  0.33 (0.015)\\
  & L2  &   0.97 (0.0050) &   0.97 (0.0050) & 0.97 (0.0050) &   0.97 (0.0050) &    0.97 (0.0050) \\
  \hline
  Second & GL & 0.99 (0.0034) &   0.97 (0.0047) &    0.92 (0.0091) &   0.81 (0.011) &    0.75 (0.012)
 \\
  half & G & 0.99 (0.0035)&   0.97 (0.0053) &    0.88 (0.010) &   0.75 (0.013) &   0.65 (0.015)\\
  & L1 & 0.93 (0.0086) &   0.93 (0.0086) & 0.93 (0.0086) &   0.93 (0.0086) & 0.93 (0.0086)\\
  & L2  &  0.97 (0.0050) &   0.88 (0.0096) &    0.52 (0.017) &    0.35 (0.016) &    0.33 (0.015)\\
  \hline
\end{tabular}
\end{flushleft}
\label{supp-performance}
\end{table}